\documentclass{aa}
\usepackage{graphicx,fixltx2e,hyperref}
\usepackage{txfonts,color}

\usepackage{natbib} 
\bibpunct{(}{)}{;}{a}{}{,}

\begin{document}

 \title{The energy of waves in the photosphere and lower
   chromosphere: IV.~Inversion results of \ion{Ca}{ii} H spectra}

   \author{C. Beck\inst{1,2} \and R. Rezaei\inst{3} \and K.G. Puschmann\inst{4}}
        
   \titlerunning{The energy of waves: IV.~Inversion results of \ion{Ca}{ii} H spectra}
  \authorrunning{C. Beck, R. Rezaei, K.G. Puschmann}  
\institute{Instituto de Astrof\'isica de Canarias (IAC)
     \and Departamento de Astrof\'isica, Universidad de La Laguna (ULL)
         \and Kiepenheuer-Institut f\"ur Sonnenphysik (KIS)
     \and Leibniz-Institut f\"ur Astrophysik Potsdam (AIP)
     }
 
\date{Received xxx; accepted xxx}

\abstract{Most semi-empirical static 1D models of the solar atmosphere in the
  magnetically quiet Sun (QS) predict a rise of the temperature at
  chromospheric layers. Numerical simulations of the solar
  chromosphere with a variable degree of sophistication, i.e., from
  one-dimensional (1D) to 3D simulations, and assuming local thermal
  equilibrium (LTE) or non-LTE (NLTE), on the other hand only yielded an increase of the brightness temperature without any stationary
  increase of the gas temperature.}{We investigate the thermal structure in
  the solar chromosphere as derived from an LTE inversion of observed
  \ion{Ca}{ii} H spectra in QS and active regions (ARs).}{We applied an
  inversion strategy based on the SIR code to \ion{Ca}{ii} H spectra to obtain
  1D temperature stratifications. We investigate the
  obtained temperature stratifications on differences between magnetic and
  field-free regions in the QS, and on differences between QS and ARs. We
  determine the energy content of individual calcium bright grains (BGs) as specific candidates of chromospheric heating events. We compare observed with synthetic NLTE spectra to estimate the significance of the LTE inversion results.}{The fluctuations of observed intensities yield a variable temperature structure with spatio-temporal rms fluctuations below 100\,K in the photosphere and 200\,--\,300\,K in the QS chromosphere. The average temperature stratification in the QS does not exhibit a clear chromospheric temperature rise, opposite to the AR case. We find a characteristic energy content of about 7$\,\times\,10^{18}$\,J for BGs that repeat with a cadence of about 160\,secs. The precursors of BGs have a vertical extent of about 200\,km and a horizontal extent of about 1\,Mm. The direct comparison of observed with synthetic NLTE profiles partly confirms the results of the LTE inversion that the solar chromosphere in the QS oscillates between an atmosphere in radiative equilibrium and one with a moderate chromospheric temperature rise. Two-dimensional $x-z$ temperature maps exhibit nearly horizontal canopy-like structures with a few Mm extent around photospheric magnetic field concentrations at a height of about 600\,km.}{The large difference between QS regions and ARs, and the better match of AR and non-LTE reference spectra suggest that magnetic heating processes are more important than commonly assumed. The temperature fluctuations in QS derived by the LTE inversion do not suffice  on average to maintain a stationary chromospheric temperature rise. The spatially {\em and\rm} vertically resolved information on the temperature structure allows one to investigate in detail the topology and evolution of the thermal structure in the lower solar atmosphere.} 
\maketitle
\section{Introduction}
The thermal structure of the solar chromosphere remains enigmatic and to the greatest extent unexplained more than a century after the first description of the existence of the chromosphere itself \citep{secchi1860,lockyer1868}. Nowadays, two competing views on the thermal structure exist, i.e., a static/stationary and a dynamical view. 

The static view is expressed by a long series of one-dimensional (1D) thermal stratifications that reproduce observed spectra \citep[e.g.,][see \citet{rutten+uitenbroek2012} and \citet{rutten2012} for additional references]{gingerich+etal1971,holweger+mueller1974,vernazza+etal1981,fontenla+etal1993,fontenla+etal2006}.
To produce emission in the line cores of chromospheric spectral lines, all
these 1D models require a reversal of the temperature gradient with height 
at some point. Their common feature therefore is the existence of a temperature
minimum at a height of about 500\,km above the photosphere, where the {\em
  gas} temperature starts to rise towards higher layers
\citep[e.g.,][]{vernazza+etal1976,ulmschneider+etal1978,leenarts+etal2011}.
Such 1D models with a temperature reversal faithfully reproduce observed
(average) spectra to the most minute level, but all lack something: they
provide a temperature stratification with height in the solar atmosphere, but do not explain the reason for its shape.

The dynamical view of the solar chromosphere is comparably newer than the
static 1D models and resulted thanks to the increase in the available
computing power during the last two decades. Opposite to the static
models that were derived from observations, the dynamical approach usually
goes the opposite way and tries to model the solar chromosphere from basic
principles, i.e., numerical simulations that encompass the
differential equations that govern the temporal evolution of material in the
solar atmosphere \citep{nordlund1982}. One of the advantages of such a forward
modeling is that one only needs to decide on what are the relevant physical processes that must be included in the set of differential equations. Whereas numerical simulations of the solar photosphere abound \citep[e.g.,][]{nordlund+stein1990,steiner+etal1998,stein+nordlund1998,steiner+etal2008,fabbian+etal2010,beeck+etal2012}, there are relatively few chromospheric simulations \citep{rammacher+ulmschneider1992,carlsson+stein1997,wedemeyer+etal2004,schaffenberger+etal2006,cuntz+etal2007,leenaarts+etal2009}. 

One reason is the spatial, and especially the vertical extent of the
simulation box that has to reach up to a height of about 2\,Mm with a sufficiently fine sampling on the order of a few tens of km to encompass the solar chromosphere. A second limitation is caused by the physical conditions in the upper solar atmosphere. The gas density in the solar photosphere is high enough that the assumption of local thermal equilibrium (LTE) is justified. In LTE, the available total energy is distributed over all available degrees of freedom by frequent collisions between particles such that each degree of freedom obtains a fully determined fraction of the total energy. The exact distribution solely depends on the physical laws holding in the material of interest. Assuming LTE, the values of several physical quantities such as population states, thermal motions, etc. are determined instantly as soon as a temperature value is given. Because of the decrease of the gas density with height in the
gravitationally stratified solar atmosphere, the LTE condition breaks down
when a sufficient number of collisions takes longer than the dynamical
evolution of the physical quantities that can change by, e.g., radiative
processes \citep[e.g.,][]{carlsson2007,wedemeyer+carlsson2011}. A complete treatment of this regime in a non-LTE (NLTE) formalism still exceeds the computing power of even the most recent computer clusters \citep{auer+mihalas1969,ayres+wiedemann1989,rammacher+cuntz1991,carlsson+stein1992,carlsson+stein1997,uitenbroek2000,asensio+etal2003,leenaarts+etal2007,wedemeyer+carlsson2011,carlsson+leenaarts2012,martinezsykora+etal2012}.

The dynamical numerical simulations of the solar chromosphere revealed a quite distinct thermal structure than the static 1D models. The {\em brightness} temperature at chromospheric layers tended to fluctuate strongly, leading to intermittent emission in the synthesized chromospheric spectra. The gas temperature, however, did not show any temperature reversal on a temporal average, contrary to most of the static 1D models \citep[cf.][]{carlsson+stein1995,carlsson+stein1997,kalkofen2001,kalkofen2012}. Thus, the average brightness and gas temperatures  in such simulations differed significantly, while the average synthesized chromospheric line spectra corresponded to neither the average gas nor the average brightness temperature. This puzzling behaviour even raised the question how to define a ``mean'' temperature at all under this conditions \citep{rammacher+cuntz2005,uitenbroek+criscuoli2011}. This point is actually more important than it seems at a first glance because the static 1D models are based on the relation between brightness temperatures (i.e., observed spectra) and gas temperature (i.e., model output). Questioning this relation between gas and brightness temperature, or the definition of temperature therefore converts directly into questioning the results of the static 1D models that provide an excellent fit to observed average spectra \citep{kalkofen2012}. 

There are some common problems for both the static and the dynamical view of the solar chromosphere. One is the width of spectral lines synthesized from the simulations or models, respectively. Observed chromospheric spectral lines are usually found
to be significantly broader than the synthetic lines, (far) in excess of the
thermal line width for any reasonable temperature range. This behaviour shows up most prominently in off-limb observations \citep[e.g.,][]{beckers1968,beck+rezaei2011}. Part of this excess in line width can be explained by the spatial resolution of the observations that represent a spatial average over solar structures with different physical properties, and especially different macroscopic mass flows with their related Doppler shifts. But also the average of synthetic spectra from simulations with spatial grid sizes down to a few ten kms lacks the observed line width, which indicates that some physical ingredient is missing in the simulations, or that their spatial sampling -- despite being below the photon mean free-path length -- still is insufficient to capture all of the dynamical evolution of, e.g., small-scale mass flows. The line width in both static and dynamical models is therefore often adjusted ad-hoc by some additional broadening parameters \citep[cf.][and references therein]{allendeprieto+etal2004,asplund+etal2004}, usually labeled as macroturbulent and microturbulent velocities that represent spatially unresolved line-broadening mechanisms of unknown origin \citep[e.g.,][]{cram1981,rutten2003}. 

A second problem is how to reconcile the temperature stratifications with the
observations of molecular lines at chromospheric heights
\citep{ayres+testerman1981}. This problem affects more the static than the
dynamical models, because in the former the location of the temperature reversal at low heights prevents any significant formation of molecules at layers above it \citep{ayres1981,kalkofen+etal1999,ayres2002}. A ``cool'' chromospheric part could partly be included also in static models by ``postponing'' the chromospheric temperature rise to higher layers \citep{ayres2002,ayres+etal2006,fontenla+etal2007,rezaei+etal2008}, but it was imposed in some sense as a post-facto crutch. In most numerical simulations, phases with a temperature that is sufficiently low to allow the formation of molecules were usually automatically present \citep{wedemeyer+etal2004,leenarts+etal2011}.

A third problem is a discrepancy between observations in different spectral
lines that yield seemingly contradictory results. Observations with the Solar Ultraviolet Measurement of Emitted Radiation (SUMER) instrument on-board the SOlar and Heliospheric Observatory (SOHO) satellite in ultraviolet (UV) lines revealed that all these lines are in permanent and ubiquitous emission with little intrinsic variation in quiet Sun regions of the solar surface \citep{carlsson+etal1997}. All these lines are classified
as ``chromospheric''. Contrary to {this, chromospheric Ca II lines,
especially H and K in the near-UV, show large spatio-temporal fluctuations
from reversal-free absorption profiles to moderate emission in the line cores \citep{liu+smith1972,cram+dame1983,rezaei+etal2008,beck+etal2008}. While the temporal evolution is naturally explained in dynamical simulations, persistent emission is only reproduced in static 1D models with a permanent chromospheric temperature rise. 

The way to resolve this contradictory behaviour of observed spectral lines and
to reconcile the conflicting views of the static and dynamical solar
chromospheric models is not clear at present. There is one additional
information to be considered: the temperature in the solar photosphere is
about 6000\,K, while the ion and electron temperature in the solar corona exceeds one million K, where both results are undoubted. Thus, there {\em has} to exist a reversal of the temperature gradient at some height, but where exactly is unclear. 

Magnetic fields could be a possible mediator between the two conflicting
scenarios of the solar chromosphere. Firstly, magnetic fields have not been
considered neither in all of the static nor many of the dynamical
models \citep[but see also][]{martinezsykora+etal2012,khomenko+collados2012}. Secondly, in quiet Sun (QS) conditions, concentrations of magnetic flux are
theoretically expected to spread out from their confined diameter in the low
photosphere to fill the full volume in the chromosphere. This structure of
several expanding flux tubes creates automatically a dividing layer at some
chromospheric height by the so-called magnetic canopy that separates
field-free from magnetized plasma \citep{giovanelli1980,steiner+etal1986,solanki+etal1991,stepan+trujillobueno2010,nutto+etal2012,holzreuter+solanki2012}, i.e., that separates a dynamical chromospheric regime below from a stationary one above the canopy \citep[as sketched for instance in][their Fig.~22]{beck+etal2008}.  

In this contribution, we investigate the spatio-temporal behaviour of the
thermal structure in the solar chromosphere as derived from an inversion of
\ion{Ca}{ii} H spectra assuming LTE \citep[][BE13]{beck+etal2012a}. This study follows the approach of the static 1D
models by deriving a gas temperature from an observed brightness temperature,
but uses spatially resolved spectra that trace the dynamical properties of the
solar chromosphere -- as in the dynamical view of the chromosphere
\citep[see also][]{henriques2012}. We use observations of both QS and
active regions (ARs) to investigate differences between these types of solar
structures. The observations and the inversion  scheme are described in detail
in BE13, we thus refer the reader to the latter publication. We will only
highlight some observational limitations that are relevant for the current
contribution in Sect.~\ref{sect_obs}. We then concentrate on the thermal
structure of the chromosphere in QS and ARs in Sect.~\ref{results}, both on
average and spatially resolved. The findings are discussed in
Sect.~\ref{discussion}. Section \ref{conclusion} provides our
conclusions. Appendix \ref{appa} discusses the possible amount of spectral
broadening by different mechanisms acting on the observed spectra.
\section{Relevant observational details \label{sect_obs}}
We use in total five different observations taken with the POlarimetric
LIttrow Spectrograph \citep[POLIS,][]{beck+etal2005b}. POLIS
delivered\footnote{POLIS was officially de-commissioned end of 2010.} intensity
spectra in \ion{Ca}{ii} H at 396.85\,nm together with spectropolarimetric data
of the two \ion{Fe}{i} lines at 630\,nm. Two data sets were taken in QS
regions, i.e., a time series of about one-hour duration and a large-area map
\citep[cf.][BE09 and BE12 in the following]{beck+etal2009,beck+etal2012}. The
other three data sets are large-area maps of the active region NOAA 10978
\citep[cf.][]{beck+rammacher2010,bethge+etal2012}. The parameters most
relevant in the present context are the slit width of 0\farcs5, the step width
in the case of spatial scanning of also 0\farcs5, the temporal cadence of
  the time series of 21\,secs, and the spatial sampling along the slit of
0\farcs292 in the \ion{Ca}{ii} H spectra. These values provide a pixel scale of 0\farcs5 $\times$ 0\farcs292 in the Ca spectra. The data were
taken with real-time seeing correction by the Kiepenheuer-Institut adaptive
optics system \citep{vdluehe+etal2003} and have a seeing-limited spatial
resolution of about 1$^{\prime\prime}$, which roughly coincides with the
sampling-limited resolution in the scanning direction.

Besides from the study of individual events in the time series (Sect.~\ref{sect_indi}), always the average of the statistics in all individual maps is used to expand the statistical base.
\begin{figure}
\begin{minipage}{8.8cm}
\resizebox{8.8cm}{!}{\includegraphics{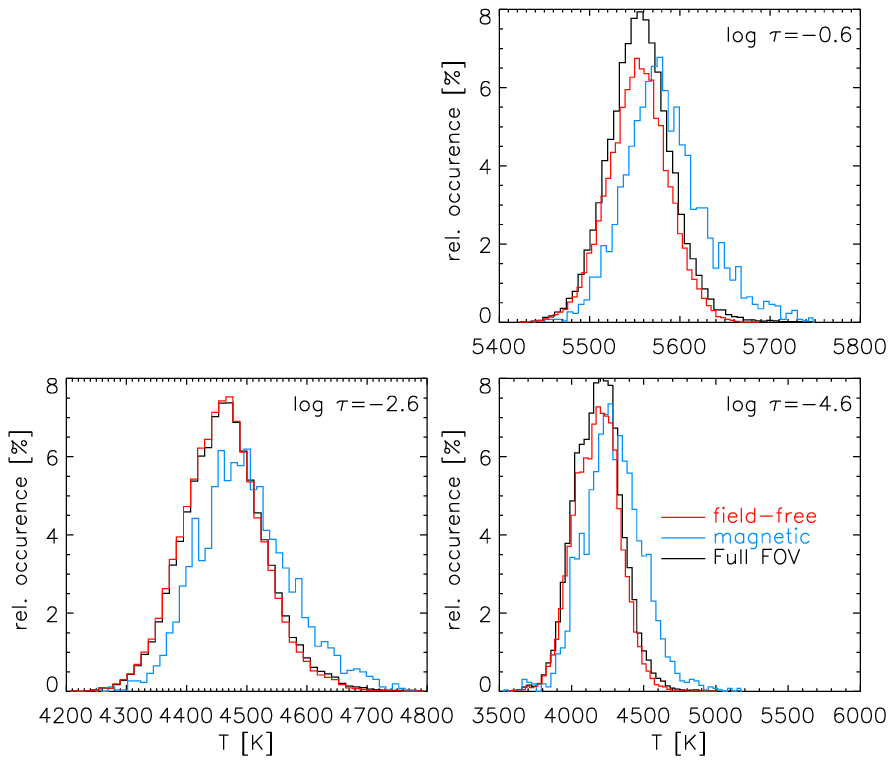}}$ $\\$ $\\
\end{minipage}\hspace*{-8.8cm}
\begin{minipage}{4.4cm}
\vspace*{-4.cm}
\resizebox{3.75cm}{!}{\hspace*{.75cm}\includegraphics{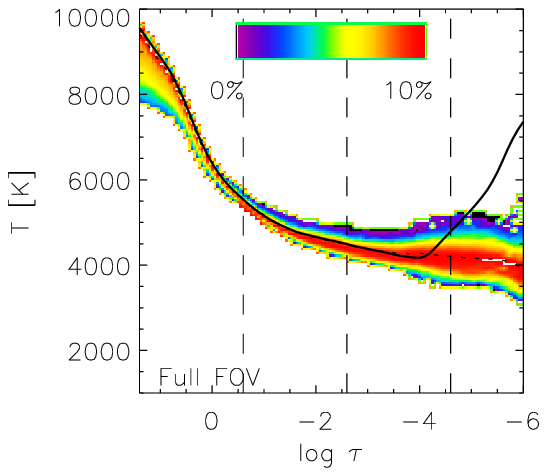}}$ $\\
\end{minipage}
\caption{Relative occurrence of temperatures in the QS. {\em Left top}: temperature versus (vs.) optical depth for the full FOV. The relative occurrence of a given temperature is indicated by the colour with the corresponding colour bar at top. {\em Thick black line}: original HSRA model. {\em Dashed}: average temperature. The {\em vertical dashed lines} denote the layers of $\log\tau$ for which the temperature histograms are shown in the rest of the panels. In the histograms, {\em black/red/blue lines} denote the full FOV/magnetic/field-free locations.\label{stat_fig}}
\end{figure}
\section{Results of LTE inversion\label{results}}
The inversion setup for \ion{Ca}{ii} H described in BE13 yields temperature stratifications for spatially resolved spectra. With the inversion results, we now can investigate the statistics of the temperature distributions at different optical depth levels $\log \tau$, analogously to BE12, where the intensity distributions at different $\lambda$ and the characteristic properties of different spatial samples in the observed field of view (FOV), i.e., field-free or magnetic regions, and the full FOV, were analyzed. The distinction between field-free and magnetic sample is made using the simultaneous polarimetric observations in the 630\,nm channel of POLIS by setting a threshold in the observed polarization signal (cf.~BE12).
\subsection{Average temperature and temperature statistics in QS\label{temp_qs}}
Following the approach of BE12, we calculated the characteristics of the temperature distributions up to second order (mean, root-mean-squared (rms) fluctuation, skewness). Figure \ref{stat_fig} shows some of the statistics of temperature versus (vs.) optical depth for the maps taken in the QS. The {\em upper left panel} shows the relative occurrence (denoted by the colour bar) of temperatures in the full FOV for each optical depth level. The three samples in the FOV (field-free, magnetic, full FOV) differed only slightly in this way of presentation, therefore only the latter is shown. The remaining panels show the histograms of temperature for all spatial samples at three optical depth levels of $\log \tau = -0.6, -2.6$, and $-4.6$, respectively. The magnetic locations ({\em blue lines}) show a weak preference for slightly higher temperatures compared to the full FOV or the field-free locations. The average temperatures of all three samples do not show an indication of a chromospheric temperature rise as predicted, e.g., by the original Harvard Smithsonian Reference Atmosphere  model \citep[HSRA,][]{gingerich+etal1971}, e.g., compare the {\em black solid line} (HSRA) and {\em black dotted line} ($<T>$ in the full FOV) in the {\em upper left panel} of Fig.~\ref{stat_fig}. 

The {\em lower panel} of Fig.~\ref{rms_fig} shows the variation of the rms fluctuations of temperature with optical depth and geometrical height. For the conversion from one to the other we used the tabulated relation given in the HSRA atmosphere model, opposite to a derivation of the height scale from the inversion results themselves as in \citet{puschmann+etal2005,puschmann+etal2010}. We note that in the derivation of the temperature statistics always the individual temperature stratifications vs.~optical depth were used, and that the rms value will (slightly) underestimate the characteristic range of fluctuations in case of asymmetric distributions (cf.~Fig.~\ref{stat_fig}). The geometrical heights in Fig.~\ref{rms_fig} only indicate a probable value for the average location of the formation heights. As in the analysis of the intensity statistics of \ion{Ca}{ii} H and \ion{Ca}{ii} IR spectra in BE12, we find that the rms fluctuations increase nearly monotonically in geometrical height, with a minimum of fluctuations near $z \sim 100$\,km (corresponding to $\log \tau \sim -1$) caused by the temperature inversion of granules and inter-granular lanes \citep[cf.][]{puschmann+etal2003,puschmann+etal2005}. The fluctuations stay below 100\,K up to 500\,km height, and reach only 200\,--\,300\,K at 1000\,km. The magnetic regions show slightly higher rms fluctuations at all heights and pass through the minimum of rms fluctuations lower in the atmosphere than the other two samples. The latter behaviour is presumably caused by the shift of the optical depth scale on magnetic locations. The maximal temperature variations ({\em dotted lines} in the {\em lower panel} of Fig.~\ref{rms_fig}) reach about 1000\,K at $\log \tau \sim -4$ ($z\sim 700$\,km).
\begin{figure}
\centerline{\includegraphics{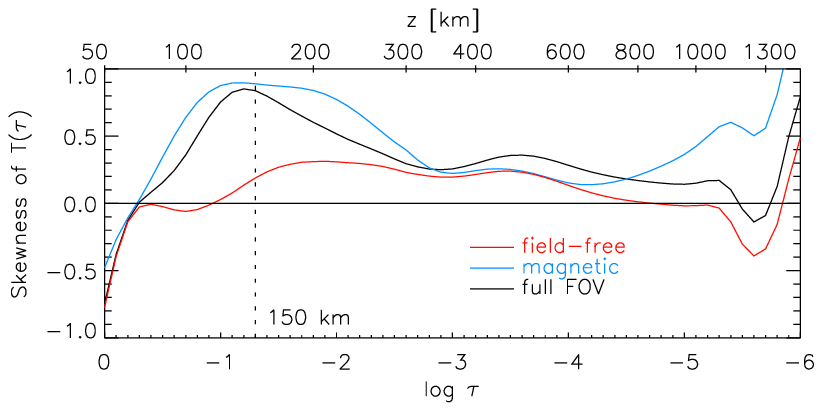}}\vspace*{.25cm}
\centerline{\includegraphics{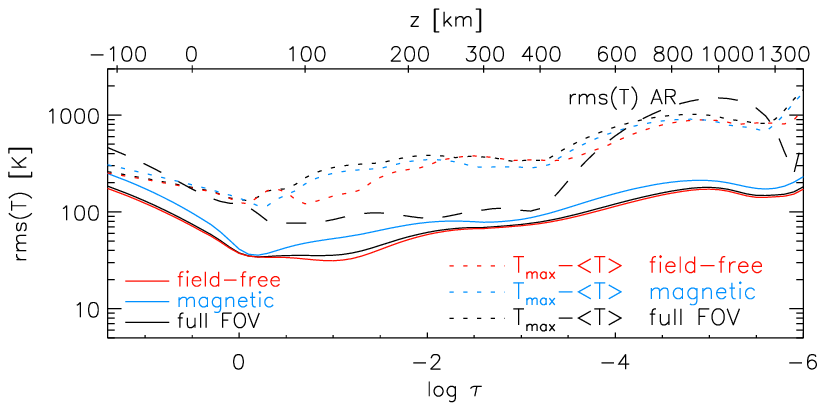}}
\caption{{\em Bottom panel}: rms fluctuations of temperature for the full FOV ({\em black}), field-free locations ({\em red}), and magnetic locations ({\em blue}). The three {\em dotted lines} indicate the maximum deviation from the average temperature. The {\em dashed line} denotes the rms fluctuation  in an AR map. {\em Top panel}: skewness of the temperature distributions. The {\em vertical dotted line} indicates $z = 150$\,km.\label{rms_fig}}
\end{figure}

BE12 found a pronounced difference in the skewness of intensity distributions between field-free and magnetic locations, i.e., an increased skewness on magnetic locations at intermediate wavelengths between the line wing and the core, corresponding to formation heights of about $z\,=\,$135\,--\,150\,km. The {\em upper panel} of Fig.~\ref{rms_fig} shows that this also happens at about the same height in the case of the temperature retrieved by the inversion. A positive skewness indicates that temperature excursions towards values above the average are more frequent than those towards values below it. The skewness on magnetic locations increases strongly already at lower optical depths of about $\log \tau \sim -4$ when compared with the other two samples. This could again be due to a shift of the optical depth scale on magnetic locations, but could also indicate that propagating waves turn into shock fronts at lower heights in the rarefied medium of magnetic flux concentrations than in the field-free surroundings. 
\begin{figure}
\centerline{\resizebox{8.8cm}{!}{\hspace*{.5cm}\includegraphics{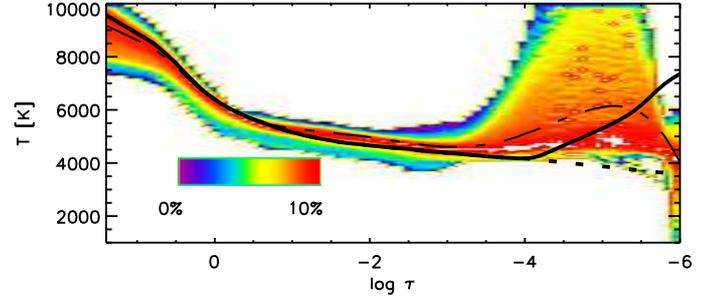}}}$ $\\
\caption{Relative occurrence of temperatures in an AR. {\em Dash-dotted}: average temperature in the AR. {\em Dotted}: average temperature in QS. {\em Thick black line}: original HSRA model.\label{ar_temp_stat}}
\end{figure}
\subsection{Average temperature and temperature statistics in AR\label{temp_ar}}
One could argue that with the list of assumptions used in the inversion (LTE,
unmodified HSRA density, lack of hydrostatic equilibrium, complete frequency
redistribution), it would be impossible to obtain a chromospheric temperature
rise in the inversion of spectra because the (partial) decoupling between
temperature and emitted intensity that exists in NLTE would be needed. Figure
\ref{ar_temp_stat} shows that this is not the case, but that a chromospheric
temperature rise results also in LTE if input spectra different to those in QS
are analyzed. Both the average temperature and the distribution of
temperatures in the AR data cover the temperature rise in the original HSRA
model (Fig.~\ref{ar_temp_stat}). The offset of the temperature stratifications
in the optical depth scale -- the HSRA temperature starts to rise at a larger
height ($\log\tau = -4$) than the average AR temperature retrieved from the
inversion ($\log\tau \sim -3$) -- is presumably caused by the off-centre
location of the AR ($\theta =50^\circ$) and the related shift of the optical
depth scale because of the inclined line of sight (LOS). We remark that the
drop of temperature at the uppermost optical depth levels ($\log\tau < -5.5$)
is purely artificial. It results from the need of having a reduction of
temperature at the uppermost atmosphere layers for generating double reversals
in the Ca line core when using the LTE approximation. The rms fluctuation of
the temperature in one of the AR maps is over-plotted in the {\em bottom panel}
of Fig.~\ref{rms_fig}. It reflects the intrinsically larger spatial variation
of temperature in an AR compared to that in QS, but we note that the range
goes up to beyond 1000\,K rms, implying even larger maximal
variations. Temperature fluctuations of this size are large enough to
  rise the temperature from an atmosphere in radiative equilibrium with
  about 4000\,K in the chromosphere to the temperature values of commonly used atmospheric models such as the HSRA.
\subsection{Correlation between intensity and temperature}
In our inversion approach, each wavelength $\lambda$ is attributed a corresponding value in $\tau$, i.e., to adjust the intensity $I(\lambda)$, the corresponding temperature $T(\tau)$ should be varied. The iterative inversion process, however, does not only modify the temperature at the given value of $\tau$, but uses a fourth order polynomial for modifying the full temperature stratification. In addition, when the radiative transfer equation through the model atmosphere is solved in the spectral synthesis, the temperature values of some extended optical depth range contribute to the intensity at one given wavelength. Therefore, the one-to-one correlation between a given wavelength and its attributed optical depth is partially broken up. Additionally, some solar surface structures such as (vertical) magnetic fields leave a signature in intensities, and hence temperature over some extended height/optical depth range. 

To determine which layer in $\log \tau$ corresponds to the intensity at a given wavelength, we calculated the linear correlation coefficient $C$ \citep[cf.][]{beck+rammacher2010} between intensity vs.~wavelength and temperature vs.~$\log\tau$. Figure \ref{corr_fig} shows the result for the large-area scan on disc centre. The correlation coefficient between intensity and temperature is similar to the intensity contribution function (e.g., Fig.~1 of BE09) in the line wing up to about 396.65\,nm, with a single maximum of correlation between $\log \tau \approx -0.5$ to $-1.5$ that slowly moves to upper layers for wavelengths closer to the line core. Between 396.65\,nm and 396.8\,nm, the correlation coefficient shows a second, slightly weaker local maximum of correlation at about $\log \tau \approx -3.5$. In the line-core region from 396.81\,nm to 396.89\,nm, two clear local maxima of correlation exist, where the upper one above $\log\tau\approx -4$ corresponds to the intensity contribution function, whereas the origin of the lower one at $\log\tau \approx -1$ is not clear at once. 
\begin{figure}
\resizebox{8.8cm}{!}{\hspace*{.75cm}\includegraphics{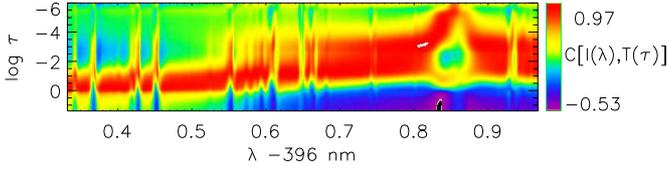}}$
$\\$ $\\
\caption{Correlation coefficient between intensity at a given wavelength
  $I(\lambda)$ and temperature $T(\log \tau)$. \label{corr_fig}}
\end{figure}

Figure \ref{corr_fig_1} reveals that the maximum of correlation between line-core intensity and temperature at low atmosphere layers is most likely caused by locations with large photospheric magnetic flux. A comparison between the line-core intensity $I_{\rm core}$ ({\em lower left panel} in Fig.~\ref{corr_fig_1}) as the most prominent tracer of photospheric magnetic flux and the temperature at optical depths of $\log\tau = -0.9, -2.5$, and $-5.4$ shows that the photospheric magnetic network can be identified at both $\log\tau = -0.9$ and $-5.4$ (correlation coefficient with $I_{\rm core}$ of 0.50 and 0.85, respectively), but is completely inconspicuous at $\log \tau = -2.5$ ($C=0.08$). This implies that vertical or only weakly inclined magnetic fields -- characteristic of the magnetic network and isolated magnetic flux concentrations because of the buoyancy force -- leave prominent signatures in temperature in at least two height ranges, i.e., at upper atmospheric layers above $\log \tau = -4$, and at lower atmospheric layers at about $\log \tau \approx -1$. This is thus most likely the reason for the increased skewness in the magnetic sample at $\log \tau \approx -1.2$ ($z\sim 150$\,km) found in the previous section and in BE12: magnetic flux concentrations and their immediate surroundings show up brighter than the quiet inter-network (IN) at intermediate wavelengths, and hence have a slightly enhanced temperature at the corresponding atmospheric layers \citep[cf.][]{sheminova+etal2005,sheminova2012}. This, however, does not explain why the intensity (temperature) is enhanced only at these wavelengths (heights), but not in between.
\begin{figure}
\resizebox{8.8cm}{!}{\hspace*{.75cm}\includegraphics{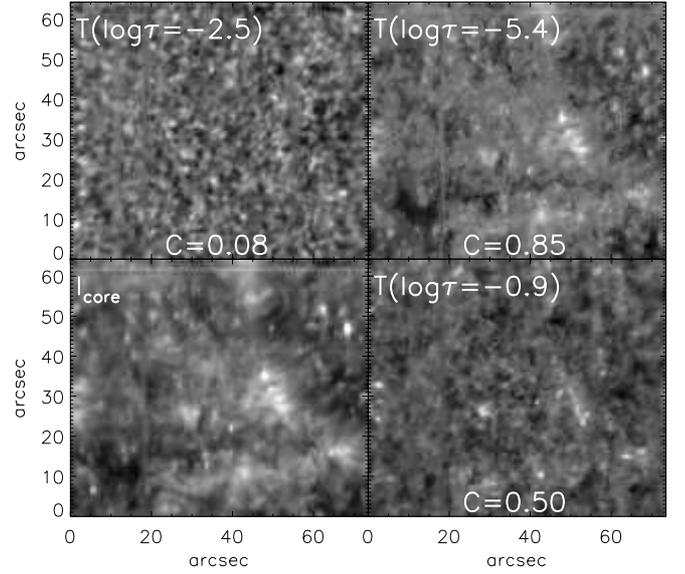}}\\$ $\\
\caption{Comparison of line-core intensity ({\em lower left}) and temperature at $\log \tau = -0.9$ ({\em lower right}), $-2.5$ ({\em upper left}), and $-5.4$ ({\em upper right}). The correlation coefficient between $I_{\rm core}$ and $T(\tau)$ is denoted in the {\em bottom middle} in the temperature panels.\label{corr_fig_1}}
\end{figure}

\subsection{Energy flux}
\subsubsection{Temperature rms fluctuations with height} 
To estimate the vertical energy flux, we determined the evolution of the energy content with height assuming that the energy balance is effected in the vertical direction
in a gravitationally stratified atmosphere with radial energy transport at
least up to the photosphere with its convective granulation. To separate the
dynamical part of the energy balance from any possible static background
caused by radiative equilibrium with the incoming photospheric continuum 
radiation, we converted the rms fluctuations of the temperature $\Delta T_{\rm
  rms}(\tau)$ -- instead of using the modulus of the temperature
$T(\tau)$ -- to the equivalent increase of internal energy using
\begin{equation}
\Delta E = \rho_{\rm gas} \cdot \frac{R \Delta T}{\mu (\gamma-1)} \;, \label{eq2}
\end{equation}
where $R = 8.31\,{\rm J\,mol}^{-1}{\rm K}^{-1}$, $\mu =  1.3\,{\rm g\,mol}^{-1}$, and $\gamma = 5/3$ are the gas constant, the specific mass, and the adiabatic coefficient, respectively (cf.~Eq.~(4) of BE09). The tabulated values of the gas density $\rho_{\rm gas}$ in the HSRA atmosphere model were used. With the relation between $\tau$ and geometrical height $z$ tabulated in the HSRA model, this yields the (stationary) average variation of energy with height ({\em top panel} of Fig.~\ref{ener_comp}).  
\begin{figure}
\resizebox{8.8cm}{!}{\includegraphics{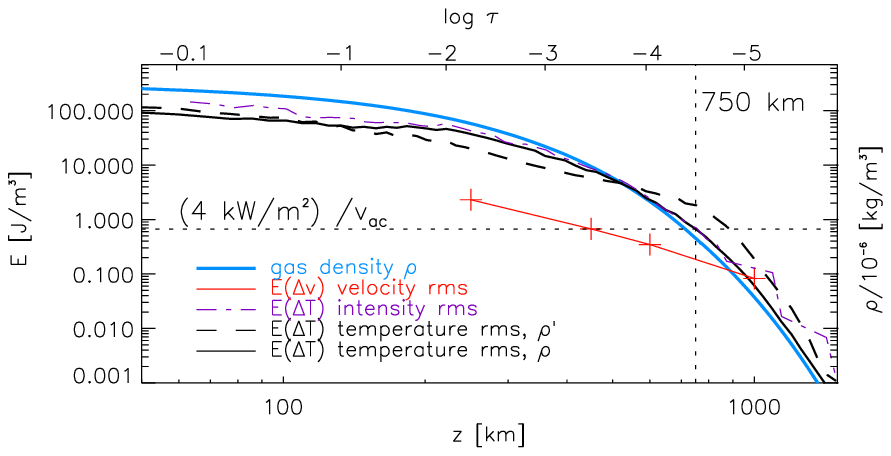}}\\
\resizebox{8.8cm}{!}{\includegraphics{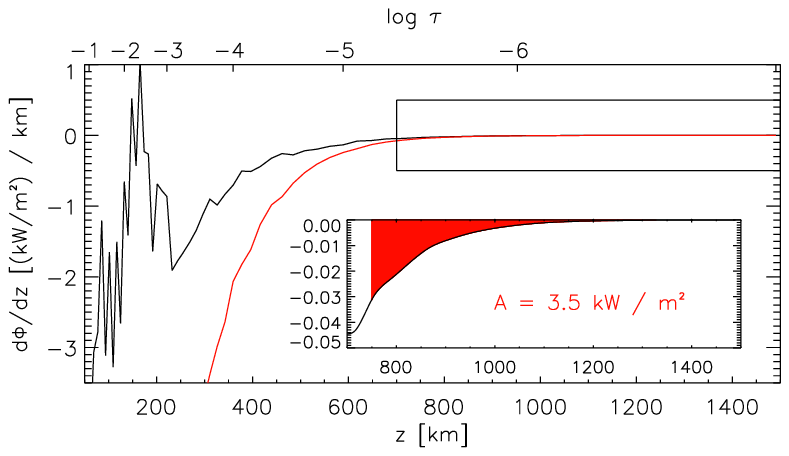}}\\
\resizebox{8.8cm}{!}{\includegraphics{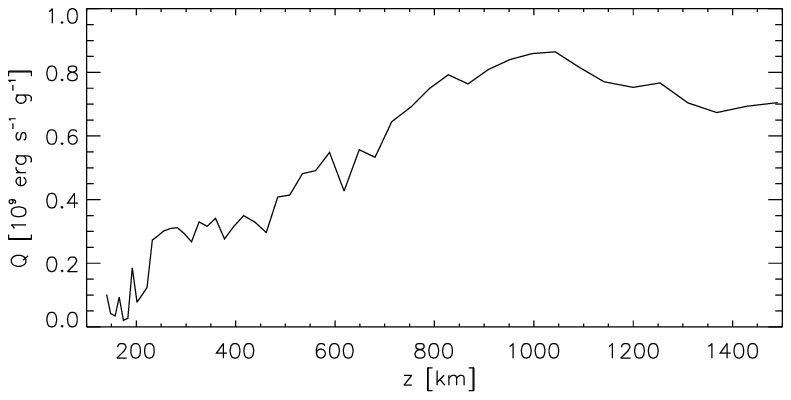}}
\caption{Variation of energy density with height. {\em Top}: energy density equivalents corresponding to velocity ({\em red line} with {\em pluses}), intensity ({\em purple dash-dotted}), and temperature rms fluctuations  ({\em black solid}). The {\em black dashed line} results for the temperature rms fluctuations when using the gas density in the original HSRA model. The {\em blue line} indicates the gas density in the modified HSRA model. {\em Middle}: energy gradient per km in height assuming a propagation with $v_{ac}$. The {\em red line} results from using a net cooling rate of 100 W\,g$^{-1}$ (CL12). {\em Bottom}: radiative losses in the units used in CL12.\label{ener_comp}}
\end{figure}

The resulting volume energy density $E/V$ can be compared with the result of
the first two papers of this series. The energy density derived from the
direct conversion of intensity fluctuations to temperature (BE12) lies
slightly above the one derived from the temperature fluctuations in the
inversion. The reason is that in the inversion the variation of intensity is
implicitly converted to a temperature fluctuation over some optical depth
range, whereas in the direct conversion from intensity to temperature all
variation is attributed to a single optical depth point. The latter approach
therefore leads to larger rms fluctuations than in the case of considering the
radiative transport through the atmosphere. The energy density corresponding
to the velocity rms (BE09) lies below that derived from intensity fluctuations
by direct conversion or by inversion up to $z\sim 1$\,Mm, but exceeds the
others above that height. We note that all temperature fluctuations - when
using $\Delta T$ without applying Eq.~(\ref{eq2}) - from either velocity,
intensity, or the LTE inversion fall short of the energy required to lift an
atmosphere in radiative equilibrium to the HSRA or any similar atmospheric
model ($\sim$\,1000\,K needed at $z\sim 1$\,Mm, but $\Delta T$ is about 500\,K
at maximum, e.g., Fig.~\ref{rms_fig}). If one assumes that the energy related
to the temperature {\em fluctuations} is transported vertically by (acoustic)
waves moving with a speed $v_{\rm ac}$ of about 6\,km\,s$^{-1}$ (cf.~Sect.~\ref{vert_speed} later on), the generic
chromospheric energy requirement of about 4\,kW\,m$^4$ of
\citet{vernazza+etal1976} corresponds to about 0.67\,J\,m$^{-3}$ ({\em
  horizontal dotted line} in the {\em top panel} of
Fig.\ref{ener_comp}). The atmospheric energy content from all three approaches
(velocity, intensity, LTE inversion) intersects with this requirement at a
height of about 750\,km, but drops rapidly below this value at higher
atmospheric layers. The ratio of the generic energy flux of
  4\,kW\,m$^{-2}$ and the energy density multiplied with a constant sound
  speed of 6\,km\,s$^{-1}$ is about ten at $z\sim 1000$\,km and increases rapidly to a factor of several hundreds at layers above.

The rapid drop of the energy density with height is caused by the
exponentially decreasing gas density $\rho_{\rm gas}$ that is over-plotted as
well in the {\em top panel} of Fig.~\ref{ener_comp}. The increase in rms
fluctuations with height is not able to compensate the decrease in $\rho_{\rm
  gas}$. We used two different gas densities for test purposes, on the one
hand the tabulated  values of the HSRA, $\rho^\prime_{\rm gas}$, and on the
other hand the gas density $\rho$ in the modified HSRA model that uses a
linear extrapolation above $\log\tau = -4$. The differences between these two
gas densities, however, do not significantly change the resulting energy
density (compare the {\em black solid} and {\em black dashed curve} in the
{\em top panel} of Fig.~\ref{ener_comp}). There is, however, a height range from about $z=100$\,km to $250$\,km, where the energy of the temperature fluctuations from the LTE inversion stays roughly constant while the density reduces by a factor of two in the same height interval, similar to the height range of constant energy found for individual bright grains (see Fig.~\ref{prop_wave_indiv} later on). 

\subsubsection{(Radiative) energy losses} 
The derivative of the energy content, again multiplied with $v_{\rm ac}$, gives the energy losses with height ({\em middle panel} of Fig.~\ref{ener_comp}). The region of roughly constant energy density can again be identified at about $z \sim 200$\,km. Integrating the area of the energy gradient for all heights above 750\,km that can be considered to contribute to the radiative losses in the Ca line core, i.e., the ``chromosphere'', yields a value of about 3.5\,kW\,m$^{-2}$, which would match the generic chromospheric energy requirement. This would, however, imply that no energy is left anymore for heating the layers above such as the corona since this amount of energy is assumed to be radiated away in all of the chromospheric spectral lines. 

For comparison with the gradient of energy losses with height, we used the
value of chromospheric radiative losses derived from a set of numerical
simulations by \citet[][CL12]{carlsson+leenaarts2012}. Their Figs.~15 and 16
yield an average cooling rate of about 100\,W\,g$^{-1}$ in the chromosphere
that can be converted to losses in energy density by a multiplication with the
gas density as a function of height. The resulting energy gradient with height
using a constant value of 100\,W\,g$^{-1}$ ({\em red line} in the {\em middle panel} of Fig.~\ref{ener_comp}) is larger than our estimate ({\em
  black line} in the {\em middle panel} of Fig.~\ref{ener_comp}) by a
factor of about 2\,--\,3 between $z\sim 200$\,--\,$300$\,km, and nearly
identical above $z\sim 800$\,km. This difference might be caused to some
extent by the fact that we only considered rms fluctuations of temperature in
the energy estimate, whereas the value of CL12 refers to all radiative losses,
i.e., losses related to the modulus of temperature itself, not its
variation. For a final comparison with CL12, we determined the energy loss
rate per mass by dividing our energy gradient with height by the mass density
({\em bottom panel} in Fig.~\ref{ener_comp}, compare with Fig.~16 of CL12). The global trend of an increase in radiative losses from the photosphere towards the chromosphere is identical in both estimates with the prediction of a maximum at about 1\,Mm. Less pronounced local maxima at $z\sim 800$ and 1300\,km appear in both estimates, albeit in the case of our estimate their significance is doubtful. In total, a roughly constant value of radiative losses of about 100\,--\,200\,W\,g$^{-1}$ in the lower chromosphere fits to both our estimate and the one of CL12.
\begin{figure}
\resizebox{8.8cm}{!}{\includegraphics{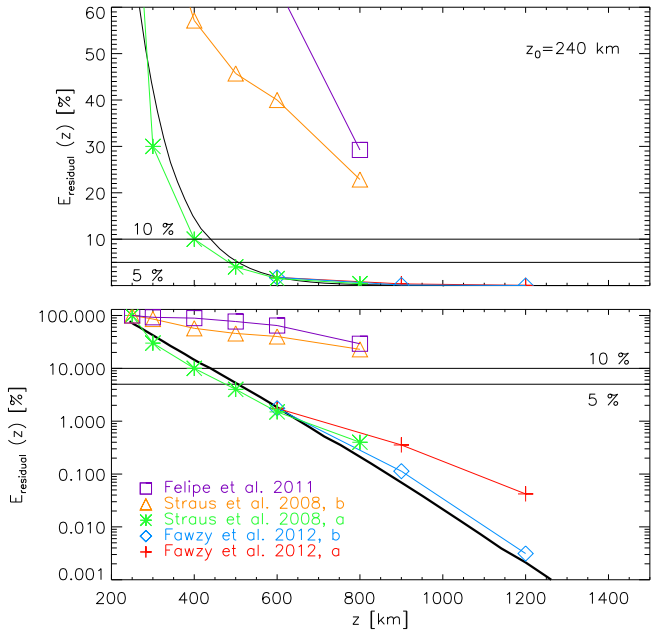}}$ $\\$ $\\
\caption{Residual of an energy input of 100\,\% at a height of 240\,km ({\em solid black line}). {\em Top}: linear scale. {\em Bottom}: logarithmic scale. The {\em red pluses} and {\em blue diamonds} give the corresponding values of FA12 for a high (case {\em a}) and low (case {\em b})  magnetic fill factor. The {\em green asterisks} give the values of ST08 for acoustic waves (case {\em a}), the {\em orange triangles} those of gravity waves (case {\em b}). The {\em purple squares} were taken from \citet{felipe+etal2011} and pertain to magneto-acoustic waves in a sunspot umbra.\label{ener_res}}
\end{figure}

\subsubsection{Residual energy content} 
Figure \ref{ener_res} shows the residual energy left at a given height when some amount of energy is dumped into the atmosphere at a height of $z= 240$\,km. To clearly visualize the effect of the energy losses with height, we provide both a linear and a logarithmic plot. The {\em black curve} corresponds to our result for the height variation of the internal energy density shown in the {\em upper left} panel of Fig.~\ref{ener_comp}, but normalized to the energy density at $z= 240$\,km. With the strong gradient of energy with height, only 5\,\% of the initial energy are left at a height of 500\,km, similar to the reduction of wave energy by 80\,\% at 400\,km found in \citet{cuntz+etal2007}. For comparison, we also overplotted the relative energy fluxes in the numerical simulations of acoustic wave propagation of \citet[][FA12, first and last column of their Table 3]{fawzy+etal2012}. Their wave energy at 600\,km was normalized to the corresponding fraction of our curve at that height. We also read off the value of the energy fluxes for acoustic waves and gravity waves from \citet[][ST08; their Fig.~3]{straus+etal2008}, both normalized to 100\,\% at $z=250$\,km. As last reference, the acoustic energy flux in a wave simulation inside a sunspot umbra from \citet[][their Fig.~16]{felipe+etal2011} is also overplotted.

A comparison of the various curves shows that our result on the variation of the energy density with height, which corresponds to the rms fluctuations of brightness temperature in the QS, matches best with all those derived for acoustic waves in an only weakly magnetized plasma (case {\em a} of ST08 and case {\em b} of FA12). This implies that any gravity waves present in the atmosphere should have no signature in intensity spectra, and hence temperature, otherwise the inversion would have had to retrieve a significantly different slope of the energy density with height as given by case {\em b} of ST08.

\subsection{Individual bright grains in the quiet Sun\label{sect_indi}}
Bright grains \citep[BGs, cf.][]{rutten+uitenbroek1991} are one of the
  most prominent features in observations of the chromospheric \ion{Ca}{ii}
lines, regardless whether two-dimensional (2D) imaging data or spectroscopic data are used. They are transient, strong intensity brightenings in chromospheric layers that appear with a random spatial pattern, but a characteristic period of about three minutes. They are clearly related to some sort of  energy
transport towards and energy increase in the solar
chromosphere
\citep[e.g.,][CS97 in the following; BE08]{carlsson+stein1997}. CS97 studied the purely hydrodynamical case of BGs in numerical simulations and were able
to show the photospheric origin of these features. 

Following the general
approach used throughout our series of papers, we will treat all BG events in
our data in the same way without trying to distinguish between a possible
acoustic or magnetic origin, which would be
possible using the spectropolarimetric data in the photospheric
\ion{Fe}{i} lines at 630\,nm. The simultaneous polarisation measurements
(cf.~BE08, their Fig.~4) showed a stable magnetic environment without clear
indications of neither magnetic flux emergence
\citep[e.g.,][]{gomory+etal2010} nor flux cancellation events \citep[e.g.,][]{bellot+beck2005}.

To estimate the energy transport related to BGs, we studied the
occurrence rate, spatial extension, and energy content of individual BGs for an
extrapolation of their total energy contribution to the chromospheric energy
balance. We used the data of the time series taken at the disc centre
(observation No.~2 in BE12) for that purpose. We define BGs as any
  temporally isolated temperature enhancement in the lower solar atmosphere,
  i.e., also below chromospheric layers.  
\subsubsection{Temporal frequency\label{sect_occurence}}
To derive the characteristic occurrence frequency of BGs, we analyzed the
temporal evolution of the temperature on the existence of local extrema, i.e.,
local temperature maxima. The {\em top panel} of Fig.~\ref{fig_eventcount}
shows the temporal evolution of the temperature excursions around the mean
value at $\log\tau = -3.6$ for one pixel along the slit as an example. The temporal evolution of the temperature is similar to the observed intensity at $\lambda = 396.83$\,nm, which roughly forms at that optical depth ({\em red dash-dotted line}). We then applied a method developed for finding polarisation lobes or local intensity reversals in line profiles \citep[e.g.,][]{beck2006,rezaei+etal2008} to all pixels along the slit and all layers in $\log \tau$ to find the locations of local maxima in the temperature excursions. To be counted as a local maxima, the value at a given instant of time has to be larger than the values on the two time steps before and after, i.e., it has to be the largest value in a range of $\pm 40\,$secs. With this approach, the influence of noise on the determination is reduced. Given the typical duration and cadence of BGs (cf.~Figs.~\ref{prop_wave_indiv} and \ref{speed_fig} later on), $\pm 40$\,secs is a characteristic time span in which usually only a single BG happens. The temperature maps were filtered beforehand for all variations with periods above 7.5\,mins to suppress the long-term variation caused by the granular evolution. We implicitly identify in the following local temperature maxima with BGs.

\begin{figure}
\resizebox{8.8cm}{!}{\includegraphics{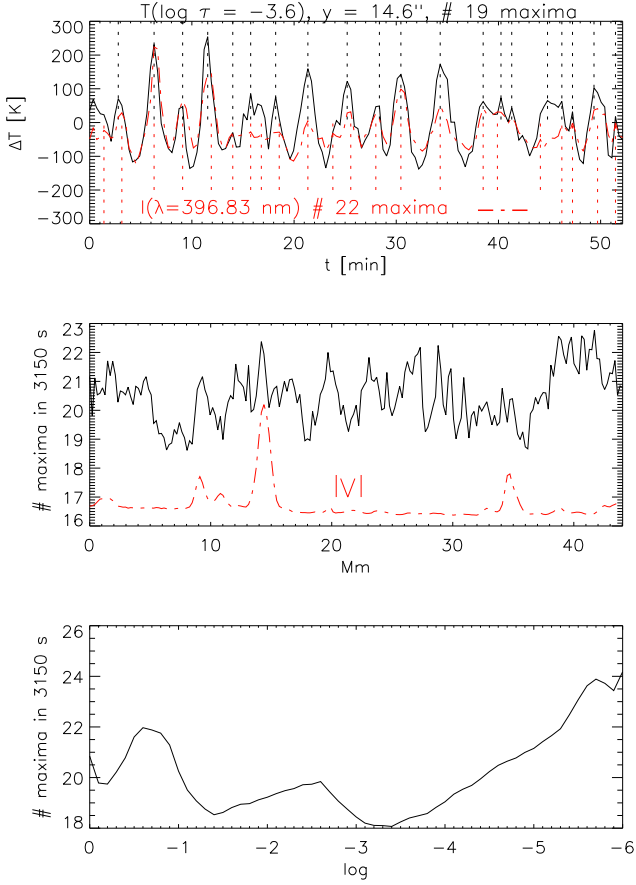}}
\caption{Temporal occurrence frequency of BG events. {\em Top}: temporal evolution of temperature at $\log \tau = -3.6$ for one pixel along the slit. The {\em red dash-dotted line} shows the intensity at $\lambda = 396.83$\,nm in arbitrary units for comparison. The {\em vertical dashed lines} indicate local maxima. {\em Middle}: temporal average of number of maxima along the slit. The {\em dash-dotted red line} shows the unsigned integrated Stokes $V$ signal in arbitrary units. {\em Bottom}: average number of local maxima during 3150\,secs as a function of $\log \tau$.\label{fig_eventcount}}
\end{figure}

Along the slit, the number of maxima was comparable for all pixels, regardless
if they were related to photospheric magnetic fields or not ({\em middle
  panel} of Fig.~\ref{fig_eventcount}). We find only a weak dependence of the
total number of maxima in the 3150 secs duration of the time series on
$\log\tau$, with a slight increase in the occurrence rate at upper atmospheric
layers ({\em bottom panel} of Fig.~\ref{fig_eventcount}). On average, we find
about 20 maxima in temperature excursions at all slit positions and layers of
$\log\tau$, which yields a characteristic cadence of about 160 seconds
\citep[cf.][]{vonuexkuell+kneer1995}. The lack of variation with optical depth
may seem puzzling at a first glance, given that usually solar oscillations are
separated into photospheric 5-min and chromospheric 3-min oscillations, but
ascribing any period to not necessarily repeated events can also be
misleading. For instance, the three BGs traced in Figs.~\ref{prop_wave_indiv}
and \ref{speed_fig} later on stay three events throughout all optical depth
layers \citep[see also][]{kalkofen+etal2010,stangalini+moretti2012}. 

The number of 20 events in 3150 secs yields an occurrence rate of about 0.006
sec$^{-1}$, which is comparable to the values given in \citet[][their Fig.~7,
0.0075 sec$^{-1}$ at maximum]{tritschler+etal2007}. We note that our rate
  is an upper limit for the occurrence of BGs because the simple 
  identification method by local temperature maxima presumably has also counted
  some peaks that could result rather from observational effects such as
  residual image jitter than from physical processes in the solar atmosphere (cf.~the {\em top panel} of Fig.~\ref{fig_eventcount} at $t\sim 40$\,min). We repeated the determination of locations of local temperature maxima with a width of $\pm 4$ time steps ($\equiv \pm 80$\,secs) or after applying a running mean of five time steps width, which both yielded only about 15 maxima on average. The dependence of the number of maxima on optical depth did not change by the additional requirement/smoothing. In case of analysing the intensity variation with time for local maxima in a $\pm 40$\,secs range ({\em top panel} of Fig.~\ref{fig_eventcount}), three additional local maxima are found in comparison to the temperature curve. Thus, the most probable range for the number of temporally local intensity or temperature maxima, and hence BGs is from 15 to 22 in 3150\,secs (0.005\,--\,0.007\,sec$^{-1}$).

The area fraction of BGs, i.e., the number of pixels along the slit that exhibit a temperature maximum at a given moment of time relative to the total number of pixels, is about 10\,\%. A similar estimate can be made from Fig.~\ref{size_fig} that shows about one BG of 1\,Mm diameter in each 2D FOV of about 1\,Mm\,$\times$\,10\,Mm.
\subsubsection{Horizontal spatial extent}
To estimate the spatial extent of individual BGs, we used the fact that the
time series on disc centre (observation No.~2 in BE12) actually consists of
150 repeated scans of 4 steps with 0\farcs5 step width each, even if only the
single step of each scan that was co-spatial in 396\,nm and 630\,nm was
investigated in detail. Figure \ref{size_fig} shows the temporal evolution of
the H$_{\rm 2V}$ intensity in a 2D FOV of $1.45$\,Mm $\times 14.8$\,Mm over
651 seconds ($\equiv$ 31 repetitions of scanning the same area). The three
consecutive brightenings of, e.g., Fig.~\ref{prop_wave_indiv} are located at
$(x,y) \approx (0-32, 6$)\,Mm. Several BGs can be identified throughout the
FOV, but their full extent in most cases is only covered along the slit in $y$
whereas  they are truncated in $x$ when they were not almost centred inside
the scanned area of 2$^{\prime\prime}$ extent (e.g., the BG in the fourth
repetition at $y\sim$\,6\,Mm in Fig.~\ref{size_fig}). 
\begin{figure}
\centerline{time $\rightarrow$ }
\resizebox{8.8cm}{!}{\hspace*{.7cm}\includegraphics{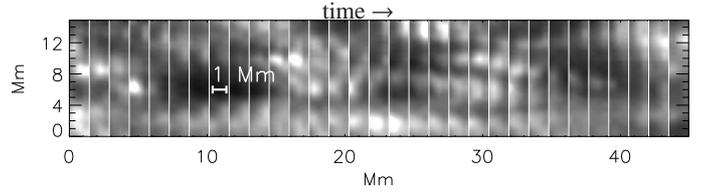}}$ $\\$ $\\
\caption{Temporal evolution of the H$_{\rm 2V}$ intensity in a 2D FOV. The {\em white vertical lines} denote the area of one scan of about 1.5\,Mm extent in $x$ taken in about 21 secs each. The {\em short white bar} at $(x,y) \approx (11,6)$ Mm indicates a length of 1\,Mm. \label{size_fig}}
\end{figure}

From the section of
  the FOV shown in Fig.~\ref{size_fig}, a roundish shape with an average
  diameter of about 1\,Mm can be attributed to BGs. To improve the statistics,
we manually identified about 100 BGs in the full FOV of all the time
series. The BGs were required to be located outside of regions with strong
polarisation signal and to be visible in at least three repetitions of
scanning the same area. We then selected the BGs at the time of their largest
brightness and took intensity cuts along the slit through their centres. We
fit Gaussians to the intensity cuts to obtain the diameters of the BGs. The
diameters ranged between 0.50 and 1.83\,Mm, with an average diameter of
1.17\,Mm. A set of about 30  BGs that happened on locations with significant
polarisation signal showed a slightly larger average diameter of 1.36\,Mm. In the following we will thus assume a typical radius of 0.5\,Mm for a single BG, which agrees well with the characteristic diameter of BGs of 1\,--\,2$^{\prime\prime}$ found in \ion{Ca}{ii} imaging data \citep[e.g.,][]{woeger+etal2006,woeger2007,tritschler+etal2007}. 
\begin{figure} 
\centerline{\hspace*{.5cm}\resizebox{5.cm}{!}{\includegraphics{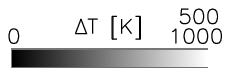}\hspace*{1cm}\includegraphics{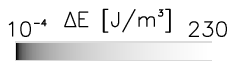}}}
$ $\\$ $\\
\centerline{\hspace*{.5cm}\resizebox{4.cm}{!}{\includegraphics{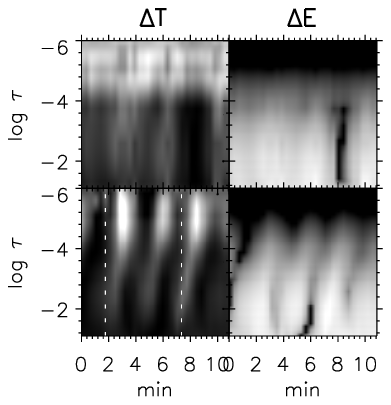}}}$ $\\$ $\\$ $\\  $ $\\  
\resizebox{8.8cm}{!}{\includegraphics{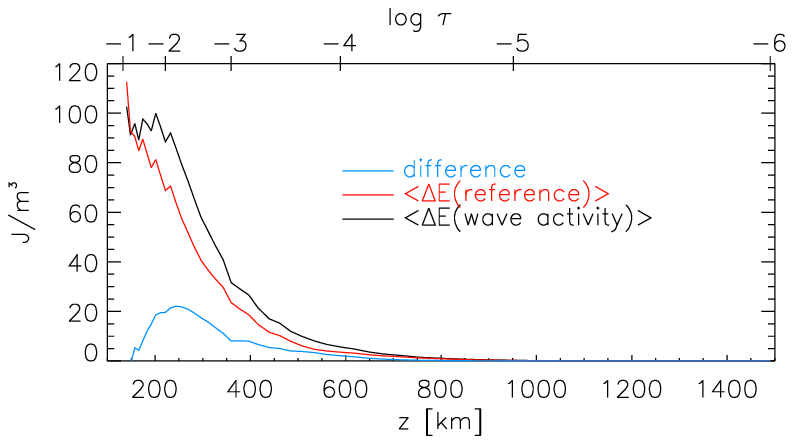}}
\caption{Estimate of the energy content of individual BG events. {\em Top panel}: temperature enhancements ({\em left column}) and internal energy enhancements ({\em right column}) for a pixel with BG activity ({\em bottom row}) and a quiet reference pixel ({\em top}). {\em Bottom panel}: average internal energy density for the active ({\em black}) and the quiet pixel ({\em red}) vs.~geometrical height. The {\em blue line} shows their difference.\label{prop_wave_indiv}}
\end{figure}
\subsubsection{Energy content}
To determine the energy content of individual BGs, one needs to derive the
temperature enhancement related to them. One additional problem here is to separate the temperature contribution of BGs from, e.g., the background (brightness) temperature that results from an atmosphere that is in radiative equilibrium with the incoming photospheric radiation. This background radiation also needs not be constant throughout the FOV because photospheric magnetic fields or variations in the granular evolution can modify it. We therefore first determined a minimum-temperature stratification $T_{\rm min}(\tau,y)$ for each pixel along the slit defined by the lowest temperature at each layer of $\log \tau$ attained on the pixel $y$ during the one-hour time series. A subtraction of this minimal temperature from the temperature stratification on the pixel $y$ at a given moment of time then yields only positive temperature fluctuations. These fluctuations represent the minimum amount of additional energy above the lowest energy state that the atmosphere exhibited during nearly one hour, without excluding that already the lowermost energetic state was energetically above radiative equilibrium. 

To isolate the BG contribution to the energy and to remove any additional
contribution from the background radiation, we first selected two different
locations for a comparison, i.e., one pixel with a significant activity in
form of a sequence of three prominent and well-defined BGs that appear
consecutively, and one pixel that showed no clear BG or other activity
(marked in Fig.~\ref{peaks_2d}). The {\em top panel} of Fig.~\ref{prop_wave_indiv} shows the temperature fluctuations relative to $T_{\rm min}(\tau,y)$ for the two pixels over a period of 651 seconds, with the ``active'' pixel in the {\em bottom row} and the ``quiet'' reference pixel in the {\em top row}. The {\em left column} shows the temperature fluctuations $\Delta T$ and the {\em right column} the corresponding variations of the internal energy density $\Delta E$ that were calculated using Eq.~(\ref{eq2}). For the conversion to geometrical height, we assumed the tabulated relation between $\log \tau$ and geometrical height $z$ of the HRSA to be valid as first order estimate also for individual spatially resolved temperature stratifications.
\begin{figure} 
\resizebox{8.8cm}{!}{\hspace*{1cm}\includegraphics{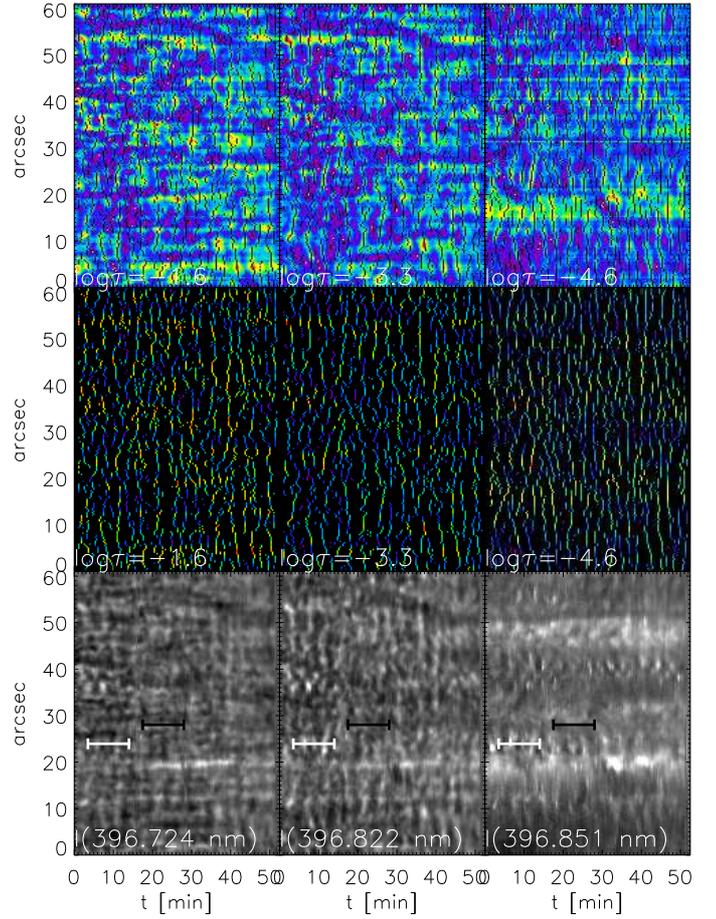}}\\$ $\\
\caption{Spatio-temporal occurrence of temperature maxima. {\em Bottom row,
    left to right}: intensity maps at $\lambda$\,=\,396.724, 396.822, and
  396.851\,nm. The {\em horizontal white/black bars} mark the locations
    of the active/reference pixel used before. {\em Middle row, left to right}: energy density at heights of $\log\tau$\,=\,$-1.6, -3.3$, and $-4.6$ at the time of temperature maxima. {\em Top}: complementary map of energy density on all other locations.\label{peaks_2d}}
\end{figure}

To determine the characteristic energy content of a BG, we averaged the value
of $\Delta E$ for the two pixels over 651\,secs ({\em bottom panel} of
Fig.~\ref{prop_wave_indiv}). The difference of the two average values of
$\Delta E$ reaches a maximum of 22.6\,J\,m$^{-3}$ at a height of about
250\,km. There are several arguments to use the difference at about this
height as the actual energy input of one BG. Around this height, the average
energy density is nearly constant for the active pixel, whereas it drops
steeply at higher atmosphere layers, indicating that the energy is inserted at
low altitudes in the atmosphere. This is also evident from the
temporal evolution of the temperature fluctuations, where the precursors of
the BG appear first low in the atmosphere in each case and then propagate
upwards. The temporal evolution of the temperature ({\em upper panel} of
Fig.~\ref{prop_wave_indiv}) also shows clearly that three BGs happen inside
the time used for the average, which are caused by three short-lived
($<\,100$\,secs) events of energy input into the low atmosphere. The temporal
fill factor -- as estimated from the time intervals when the BGs can be
  clearly identified in the temperature variation in Fig.~\ref{prop_wave_indiv} -- of all three BGs together in the time span considered is about 50\,\%. The energy enhancement of a single BG thus should correspond to about twice the temporally average value, i.e., about 45\,J\,m$^{-3}$. 

To improve the statistical base of the result and to ensure that it is not (strongly) biased by exactly the train of BGs and the reference pixel that we selected, we also calculated the statistics of BGs relative to non-BGs for the full FOV. To that extent, we used the mask of locations with local temperature maxima determined in Sect.~\ref{sect_occurence}. The internal energy density of all locations, where local temperature maxima at three different optical depths occurred, is shown in the {\em middle panel} of Fig.~\ref{peaks_2d}, whereas the {\em top panel} shows the complementary map of the energy density. The {\em bottom panel} shows the intensity at the wavelengths that were attributed to these optical depths. A comparison of the {\em middle} and {\em upper panel} reveals that the BGs with their apparent diameter of 1\,Mm are usually only a part of a wave front that extends over up to ten Mm \citep[cf.][]{vonuexkuell+kneer1995} and induces temperature maxima along its full length. The temperature maxima mark the spines of intensity brightenings, with a similar rise and decay time of the BG events ({\em top panel} of Fig.~\ref{peaks_2d}). The temporal extent appears strongly compressed in the spatio-temporal map in comparison to the spatial extent. 
\begin{figure}
\resizebox{8.8cm}{!}{\includegraphics{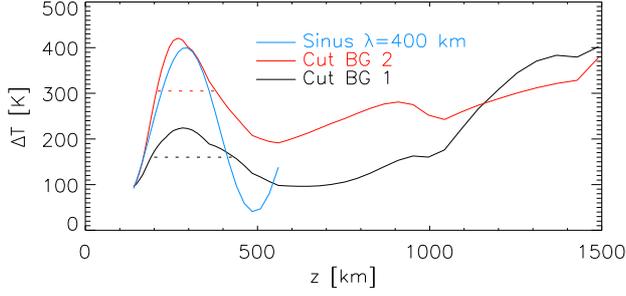}}
\caption{Temperature stratifications vs.~geometrical height at the onset of two BGs corresponding to the {\em vertical white dotted lines} in the {\em top panel} of Fig.~\ref{prop_wave_indiv}. The {\em horizontal dotted lines} denote the FWHM of the BG events. The {\em blue line} shows a sinus curve with a wavelength of 400\,km for comparison. \label{vertical}} 
\end{figure}

To determine the maximal energy enhancement caused by a BG, we calculated the
average value of the energy density for all temperature maxima and for the
complementary sample, i.e., all points in time and space that did not
correspond to a local temperature maximum. For the optical depth of $\log\tau
= -1.2$ ($\equiv 148$\,km), BGs have an average energy density of
135\,J\,m$^{-3}$, whereas the corresponding value at  $\log\tau = -2.2$
($\equiv 244$\,km) is 124\,J\,m$^{-3}$. For the non-BG sample, the average
energy density is 89\,J\,m$^{-3}$ and 80\,J\,m$^{-3}$, respectively. We
excluded not only the location of the temperature maximum, but also always the
directly preceding and following time step in the calculation of the average
non-BG energy. The energy difference between BGs and non-BGs then is
46\,J\,m$^{-3}$ and 44\,J\,m$^{-3}$ at 150\,km and 250\,km height,
respectively, matching the estimate above from individual BGs. We will use a
characteristic enhancement of the internal energy density of 45\,J\,m$^{-3}$
for one BG in the following.
\subsubsection{Vertical extent}
To determine the vertical extent of BGs, in principle the curve of the energy difference between the quiet and the active pixel could be used ({\em lower panel} of Fig.~\ref{prop_wave_indiv}), but we decided to use the temperature stratifications instead because of the larger amplitude of the temperature fluctuations. We converted the temperature stratifications at the onset of two BGs (indicated by the {\em white vertical dotted lines} in the {\em lower left panel} of Fig.~\ref{prop_wave_indiv}) from optical depth to geometrical height (Fig.~\ref{vertical}). The BGs at this stage of their evolution show up as Gaussian-like temperature bumps centred at about $z\sim 300$\,km. The full-width at half-maximum (FWHM) of the temperature enhancements is about 200\,km in height, which would correspond to half a period of a sinusoidal wave with 400\,km wavelength ({\em blue line} in Fig.~\ref{vertical}). Identifying BGs with acoustic waves with a period of 3 or 5\,min and assuming propagation with sound speed (about 6\,kms$^{-1}$) would predict significantly larger wavelengths of 1 and 1.8\,Mm instead. The propagation of BGs in the atmosphere is not purely vertical, but has also a horizontal component (cf.~Figs.~\ref{size_fig} and Fig.~\ref{peaks_2d}, some of the BGs move along the slit with time), which modifies the apparent vertical wavelength. The angle between local vertical and the wave propagation directions seems, however, to be small in most cases. 

With the information on the vertical extent and the effective enhancement of the internal energy density, one can derive an estimate of the total energy of a BG by $E = V \cdot \Delta E$. For a single pixel of 0\farcs5$\times$0\farcs292 extent, assuming a vertical extent of 200\,km, and using $\Delta E = 45$\,Jm$^{-3}$, one obtains a total energy of the BG of 6.91$\,\times\,10^{17}$\,J. Assuming a roundish structure with a radius of 0.5\,Mm, the total energy of a BG is about 7.1$\,\times\,10^{18}$\,J. 
\begin{figure}
\centerline{\resizebox{4cm}{!}{\includegraphics{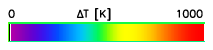}}}$ $\\
\centerline{\includegraphics{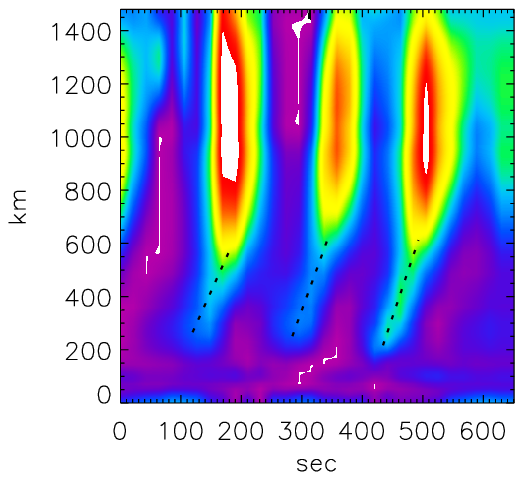}}$ $\\$ $\\
\caption{Vertical propagation speed of temperature perturbations. The background image shows the temperature variation of the active pixel of Fig.~\ref{prop_wave_indiv} vs.~geometrical height. The {\em three dotted lines} trace the propagation of the perturbations with time.\label{speed_fig}}
\end{figure}
\subsubsection{Vertical propagation speed\label{vert_speed}}
For estimating the vertical energy flux, the characteristic vertical
propagation speed of BGs is needed. The LTE inversion provides the necessary
information in form of temperature variations in optical depth. We used {again
  the tabulated relation between $\tau$ and $z$ of the HSRA model to convert
  from optical depth to a geometrical height scale. Figure \ref{speed_fig}
  shows the resulting temperature fluctuations at the active pixel
  vs.~geometrical height. The precursors of the BGs leave inclined traces in
  the lower part of the atmosphere up to about 600\,km height, whereas above
  the temperature enhancements are nearly vertical, i.e., seemingly affecting
  all higher layers at the same time. The latter presumably does not indicate
  that the propagation speed increases strongly, but rather that the relation
  between $\tau$ and $z$ is not valid inside a shock front. CS97 concluded from their 1D NLTE simulations that the emission in \ion{Ca}{ii} H at maximal brightness of a BG stems from a height range of about $z\,\sim$\,1 to 1.4\,Mm, where the strongest contributions to the emergent intensity originate in shocks. For the lower part and the onset of the BGs, the propagation speed of the temperature perturbations can, however, well be measured ({\em inclined dotted lines} in Fig.~\ref{speed_fig}). The corresponding slopes denote vertical propagation velocities of about 5.0, 6.3, and 6.7\,km\,s$^{-1}$, respectively, comparable to the sound speed \citep[see also][]{vonuexkuell+etal1983}. We note that an estimate of the propagation speed and direction in all three spatial dimensions can be derived from an inversion of all steps of each repeated scan, dropping the restriction of using only the step that was co-spatial in the \ion{Ca}{ii} H and 630\,nm spectra.

The magnified view of Fig.~\ref{speed_fig} on the BG events allows one to estimate an additional characteristic, the typical duration. Above a height of 600\,km, none of the BG events lasts longer than about 60\,secs from its onset to its subsequent complete disappearance. 

\subsubsection{Energy flux related to BGs}
The temporal evolution of the intensity or temperature shows
that the energy of BGs is removed by both propagation and radiation in a short time. There are, however, different ways to estimate the vertical energy flux per unit area. Using the energy content of a single BG of 6.91$\,\times\,10^{17}$\,J per pixel of 0\farcs5$\times$0\farcs292 as estimated above and assuming that all this energy disappears again before the onset of the next BG 160\,secs later yields an energy flux of 56.3\,kW\,m$^{-2}$, but without specifying into which spatial direction (or into which degree of freedom) the energy is dissipated or disappears. The average energy enhancement of 22.6\,J\,m$^{-3}$ of the active relative to the quiet pixel (Fig.~\ref{prop_wave_indiv}) yields an energy flux of 41.4\,kW\,m$^{-2}$ from both the influx and dissipation of the energy of three BGs in a time span of 651\,secs. Excluding a downward energy transport and otherwise equal probability for an energy transport to the remaining five possible spatial directions, the upwards vertical energy flux would be about $(41.4$\,--\,$56.3)/ 5 \sim (8.3$\,--\,$11.3)$\,kW\,m$^{-2}$. This estimate uses the characteristic temporal evolution in the derivation, i.e., about 160\,secs cadence between BGs or three events in 651\,seconds, respectively, and refers to the energy flux between about 150 to 350\,km height. 

However, even if individual BGs induce a vertical energy flux of up to a few tens of kW\,m$^{-2}$ in the low solar atmosphere ($z < 400$\,km), it seems that at $z\sim 1$\,Mm far less than 5\,\% of the energy flux are left over (Fig.~\ref{ener_res}). We remark that in our case the underlying curve that predicts this behaviour is directly proportional to the temperature rms with height and the gas density (Eq.~\ref{eq2}). Modifications in the stratification of either of the two quantities therefore will change the residual energy as well. 
\begin{figure}
\centerline{\resizebox{8.8cm}{!}{\includegraphics{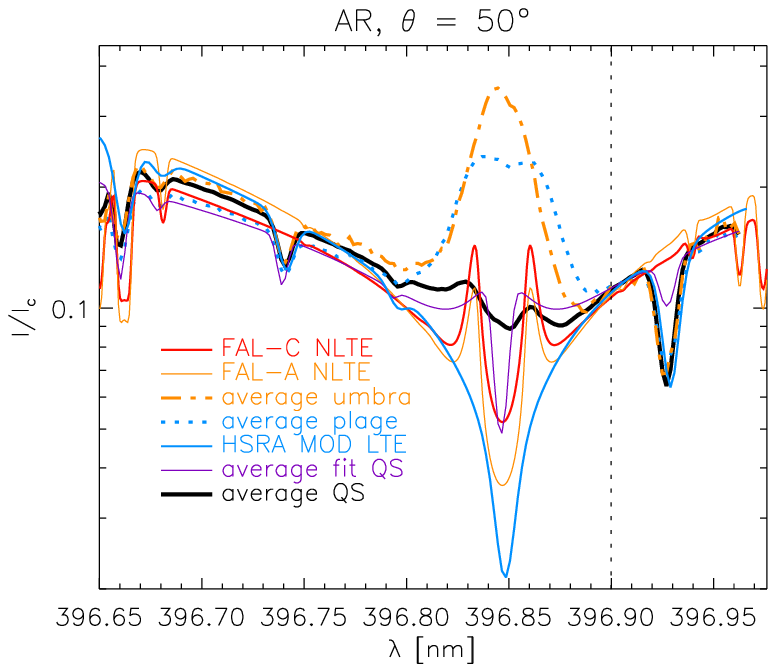}}}
\centerline{\resizebox{8.8cm}{!}{\includegraphics{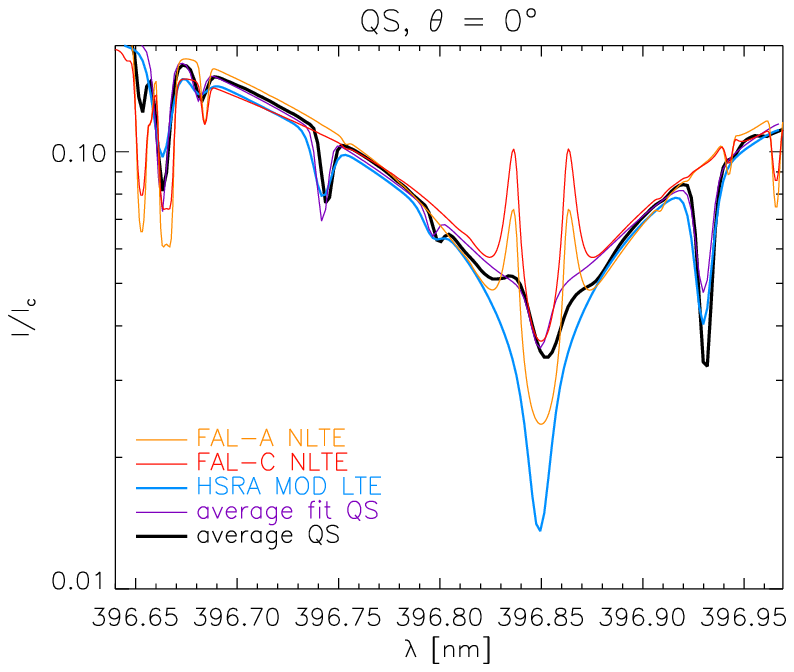}}}
\caption{Comparison of average observed and best-fit profiles with theoretical NLTE profiles. {\em Bottom panel}: spectra from a time series in the QS on disc centre. {\em Top panel}: spectra from an AR map off centre. {\em Thick black}: average observed QS profile. {\em Thin purple}: average best-fit QS profile. {\em Blue}: LTE profile of modified HSRA. {\em Red/orange}: NLTE FAL-C/A profile. For the AR, average profiles of a plage area ({\em blue dotted})  and an umbra are over-plotted ({\em orange dash-dotted}). The {\em vertical dotted line} denotes the wavelength of intensity normalization. \label{av_profs}}
\end{figure}
\section{Discussion \label{discussion}}
The reliability of our analysis results depends to some extent on the LTE assumption taken. We therefore first return to a discussion of observed and theoretical profiles that is partly independent of the (N)LTE question. 
\subsection{Comparison of average profiles and corresponding temperature stratifications}
Figure \ref{av_profs} shows the average observed and best-fit profile in QS for our time series and for one of the AR maps, an average umbral and plage profile from the same AR map, the profile synthesized in LTE from the modified HSRA model without a temperature rise, and two profiles synthesized in NLTE from the FAL-A and FAL-C temperature stratifications \citep{fontenla+etal2006} using the code of \citet{uitenbroek2001}. The spectral sampling of the FAL NLTE profiles in the synthesis was 0.75\,pm without any additional spectral degradation (cf.~the next section). To facilitate an easier comparison of the variations in the very line core in this figure, all profiles from the AR map were normalized to the same intensity at 396.9\,nm. 

Both NLTE FAL profiles exceed the observed emission peaks in the QS on disc centre ({\em lower panel} of Fig.~\ref{av_profs}). The FAL-C profile exceeds even the emission in the QS region of the AR map ({\em top panel}). To exclude a possible influence of the large-scale averaging that could reduce the amplitude of the emission peaks through the presence of spatially/temporally varying Doppler shifts, we calculated a QS profile that was averaged only over 112 spectra, i.e., one slit position in the time series. This small-scale average QS profile showed a slightly increased amplitude of the H$_{\rm 2V}$ emission peak, but the change stayed minor and the intensity still fell far short of that in the NLTE FAL profiles. Both the observed and average best-fit profile, however, clearly exceed the profile corresponding to the modified HSRA atmosphere. Average profiles in the QS therefore fall between an profile from an atmosphere in radiative equilibrium and an atmosphere with a chromospheric temperature rise such as in FAL-C or FAL-A. For the case of the AR map, the profiles change significantly. The observed average QS profile has two clear emission peaks, and the average best-fit profile in the QS starts to resemble the NLTE FAL profiles. We remark that inside the AR map, the ``QS'' regions contain far more magnetic flux than found in the QS at disc centre, and furthermore that the AR observations were taken at a heliocentric angle of about 50$^\circ$. The emission in average plage and umbra exceeds the FAL NLTE profiles by a factor of about two. 
\begin{figure}
\centerline{\resizebox{8.8cm}{!}{\includegraphics{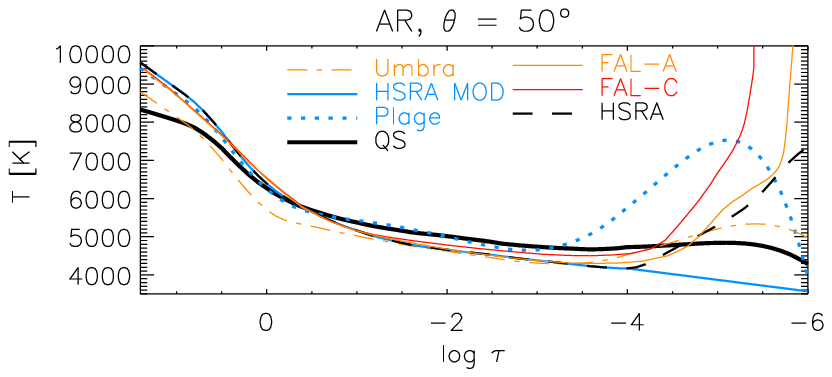}}}
\centerline{\resizebox{8.8cm}{!}{\includegraphics{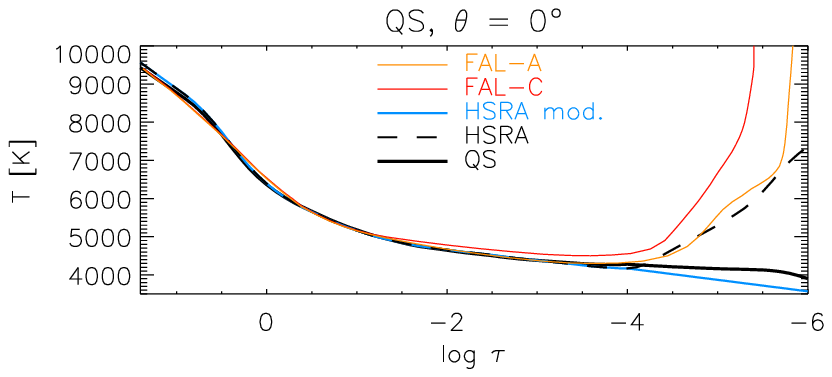}}}
\caption{Comparison of temperature stratifications. {\em Bottom}: QS maps. {\em Top}: one of the AR maps. {\em Thick black}: average QS temperature in the LTE fit. {\em Blue}: modified HSRA model. {\em Dashed}: original HSRA model. {\em Red}: FAL-C. {\em Orange}: FAL-A. For the AR, the average temperature stratification of a plage area ({\em blue dotted}) and of the umbra are over-plotted ({\em orange}) instead.\label{fig_temp_comp}}
\end{figure}

The corresponding temperature stratifications in Fig.~\ref{fig_temp_comp} reflect closely the differences in the profiles themselves. The modified HSRA lies below the  average temperature of the QS on disc centre retrieved by the inversion, and the two FAL models above it, whereas the stratifications obtained by the inversion for the AR region plage and umbra follow  to some extent the chromospheric temperature rise in the FAL models or the original HSRA model ({\em top panel} in Fig.~\ref{fig_temp_comp}). 

The LTE assumption for sure has a tremendous effect on the retrieved
temperature stratifications, but the main problem consists already in
reconciling the observed average spectra in QS with those predicted by the
NLTE FAL profiles ({\em lower panel} of Fig.~\ref{av_profs}). Appendix
  \ref{appa} discusses possible spectral broadening mechanisms that could
  reduce the amplitude of the emission peaks in observed spectra. We find that
  any reasonable instrumental broadening, or broadening by unresolved
  micro- and macroturbulent velocities with reasonable velocities ($<
  10$\,kms$^{-1}$) are insufficient to reduce the emission peaks from the NLTE
  FAL models enough to fully match the observed spectra, leaving as only
  option a clear reduction of the chromospheric temperature \citep[cf.][]{rezaei+etal2008}.
\subsection{Difference between temperature stratifications in quiet Sun and active regions}
Even if it is well known that the emission in the \ion{Ca}{ii} H line core increases in the presence of (photospheric) magnetic fields \citep[e.g.,][]{simon+leighton1964,linsky+avrett1970,liu+smith1972,dunn+zirker1973,mattig+kneer1978,schrijver+etal1989,rezaei+etal2007}, the difference between temperatures in QS and AR retrieved by the inversion is surprisingly large. The LTE approach used in the inversion retrieves temperature stratifications with a clear chromospheric rise in the AR, even though it supposedly underestimates the temperature. That the underestimation comes with a caveat is caused by a dichotomy in the (N)LTE problem. In the LTE assumption, (gas) temperature and intensity are directly coupled. In case that an increase in emission is not caused by an increase in the gas temperature but by NLTE effects on the emitted radiation field, the LTE inversion will attribute it still to a temperature change and will yield a {\em too high} gas temperature. In case that the emission is caused by an increase in the gas temperature, the LTE approach will yield a {\em too low} temperature, because it assumes that the corresponding energy will be evenly distributed over all degrees of freedom, whereas in NLTE a significant fraction can go into additional ionization or other energy sinks without increasing the emitted radiation. 

For active regions, the temperature structure in the solar atmosphere seems to
be more stable than in the QS. While, e.g., umbral profiles show clear
indications for upwards propagating wave patterns producing shock waves in the
chromosphere like in the QS
\citep[e.g.,][]{lites1986,centeno+etal2006,felipe+etal2010},  \ion{Ca}{ii} H
spectra in umbrae never loose their emission peaks
\citep{mattig+kneer1978,kneer+etal1981,kneer+etal1981a,vandervoort+etal2003,felipe+etal2010},
in contrast to QS profiles that oscillate between low or no emission and
intensity reversals
\citep[e.g,][BE12]{liu1974,cram+dame1983,rezaei+etal2008}. The strong
variability in QS was also found in numerical simulations \citep[CS97,][]{wedemeyer+etal2004,leenarts+etal2011}, but the corresponding prediction for active regions or the interior of sunspots is still pending. Even if we cannot provide any conclusive proofs, the results of our LTE inversion support a more stable gas temperature structure in active regions with a clear chromospheric temperature rise. This does not exclude that the line shape is caused by NLTE effects, but suggests that the underlying temperature stratification seems to be less influenced by temporal variations than in the QS. 

The modulus of the difference between QS and active regions even in our LTE approach indicates that heating processes related to strong magnetic fields are up to five times stronger than those related to acoustic waves in the QS: the (spatial) temperature rms in the AR is about five times larger than the spatial or temporal rms in QS (Fig.~\ref{rms_fig}) and the enhancement of the QS temperature stratification above the modified HSRA model ($\Delta T\,<\,$1000\,K, Fig.~\ref{fig_temp_comp}) is roughly five times smaller than for the average plage ($\Delta T\sim 4000$\,K). Because in most of the studies of low chromospheric lines such as \ion{Ca}{ii} H the presence of magnetic fields was neglected, their inclusion could be the turning point to reconcile the high variability of, e.g., \ion{Ca}{ii} H with the low variability of chromospheric lines observed by SUMER \citep{carlsson+etal1997}. We point out that the latter authors also found a permanent presence of emission in the SUMER lines in QS, in conflict to \ion{Ca}{ii} H spectra in QS, but in full agreement with the behaviour of \ion{Ca}{ii} H spectra in the umbra. This would fit all together with the assumption that the SUMER lines form in a magnetized atmosphere (like \ion{Ca}{ii} H in the umbra) in QS, e.g., above a magnetic canopy with a strong damping of impinging acoustic oscillation, whereas \ion{Ca}{ii} H forms below the canopy in QS \citep[cf.][their Fig.~22]{beck+etal2008}. We note also that the same effect is suggested by the numerical simulations of FA12. The average temperature stratifications shown in their Fig.~6 do not exhibit a clear temperature reversal below a height of 1\,Mm, and additionally show a strong increase of the temperature when the magnetic fill factor is increased by an order of magnitude, caused by the related reduction of the gas density in a partially evacuated magnetized plasma.
\subsection{Energy content of bright grains and related energy flux}
We estimated the energy content of the precursors of chromospheric bright grains. We find a spatial extent of about 1\,Mm by 0.2\,Mm in height, with an increase of the internal energy by about 45\,J\,m$^{-3}$ and a total energy of about 7$\,\times\,10^{18}$\,J per event. The temperature enhancements propagate upwards with about 6\,km\,s$^{-1}$, comparable to the local sound speed \citep[see also][]{worrall2012}. With the typical temporal cadence of 160\,s for individual events this yields a vertical energy flux of about 10\,kW\,m$^{-2}$ at a height of about 250\,km. Some insecurity in this calculation is related to the characteristic cadence because it is sensitive to how a local maxima is identified and counted. Our determination yields from 15 to 22 maxima in 3150\,secs, which gives roughly a 20\,\% error on the energy flux. Our specific data is less suited for the determination of the occurrence rate of BGs than time series of intrinsically two-dimensional data. The number of about 10\,kW\,m$^{-2}$ falls in line with the energy fluxes between 3 to 12\,kW\,m$^{-2}$ at about 250\,km height determined from velocity oscillations in, e.g., \citet{nazi+etal2009,nazi+etal2010} or \citet[][MA12]{malherbe+etal2012}. The latter authors also determined the total energy contained in what they call ``acoustic events'', i.e., a phase of upwards directed energy transport that can cover a few wave trains and can last up to 10\,min (cf.~Fig.~9 of MA12). They found total energies of about 1.9$\,\times\,10^{19}$\,J per acoustic event, which fits to the example used in our Sect.~\ref{sect_indi} with three BGs of in total about 2.1$\,\times\,10^{19}$\,J in 651 seconds. 

MA12 found an occurence rate of acoustic events of about 5.7$\,\times\,10^{-16}$\,s$^{-1}$\,m$^{-2}$. With 20 temperature extrema in 3150\,secs on each pixel of 0\farcs3$\,\times\,$0\farcs5, and a spatial fill fraction of BGs of about 10\,\% along the slit, we obtain a rate of 8.3$\,\times\,10^{-15}$\,s$^{-1}$\,m$^{-2}$, or about 0.5\,min$^{-1}$\,Mm$^{-2}$, respectively. The difference between the two rates is about a factor of 15, where our rate should be larger by a factor of 3\,--\,5 because of the existence of multiple temperature maxima during an acoustic event as defined by MA12. The remaining difference could be caused by the lower spatial resolution of our data, where the area of BGs will be slightly overestimated because of the spatial smearing. The recently inaugurated 1.5-m GREGOR telescope \citep[][]{schmidt+etal2012} will offer the possibility to obtain high-resolution, high-cadence spectra of the \ion{Ca}{ii} IR lines around 850\,nm with the Gregor Fabry-Per\'ot Interferometer \citep[GFPI, e.g.,][]{puschmann+etal2012c,puschmann+etal2012a,puschmann+etal2012b} and of \ion{Ca}{ii} H with the BLue Imaging Solar Spectrometer \citep[BLISS,][]{puschmann+etal2012b} in the near future.
\begin{figure*}
\vspace*{-.7cm}
\centerline{\resizebox{6.5cm}{!}{\includegraphics{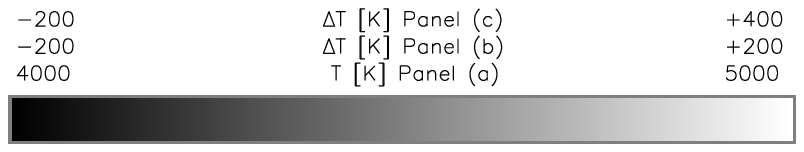}}}\vspace*{.0cm}
\resizebox{17.6cm}{!}{\hspace*{1cm}\includegraphics{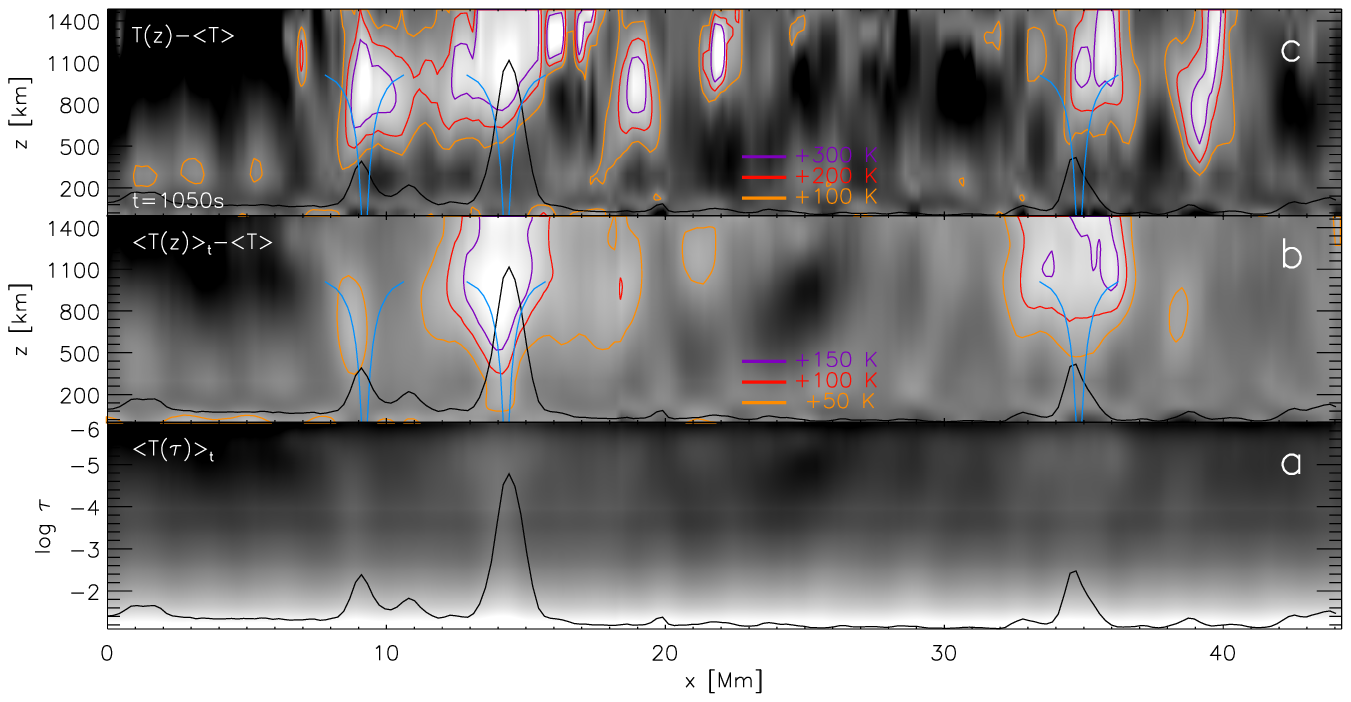}}\\$ $\\
\caption{Spatial temperature and velocity maps of the time series. {\em Panel
    a}: temporally averaged temperature $<$$T$$>_t$ vs.~optical depth. {\em
    Panel b}: $<$$T$$>_t$ minus temporally {\em and} spatially averaged
  temperature stratification $<$$T$$>$ vs.~geometrical height. {\em Panel c}:
  temperature at one scan step minus $<$$T$$>$ vs.~geometrical height. 
    The $z$-axis in the {\em upper two panels} is stretched by a factor of
    five relative to the $x$-axis. The {\em black lines} at the lower border of each panel denote the total circular polarisation along the slit in arbitrary units. The {\em blue lines} show one flux tube model of SO91. The {\em orange/red/purple} contours in panel {\em b} ({\em c}) mark temperature excesses of 50, 100, 150\,K (100, 200, 300\,K).\label{previous_fig}} 
\end{figure*}

Photospheric energy fluxes in the range of a few kW\,m$^{-2}$ would suffice to replenish the generic chromospheric energy requirement of 4\,kW\,m$^{-2}$, if -- and only if -- this amount of energy would finally reach the chromosphere. The estimates of the energy losses with height, however, predict that less than 5\,\% of an energy input at 250\,km height reaches layers above 400\,km. This suggests that the chromospheric radiative losses cannot be balanced by the energy of propagating waves or temperature perturbations despite a sufficiently large reservoir of mechanical energy in the photosphere \citep[but see also][]{petrukhin+etal2012}. We note that the temperature stratifications and the rms temperature fluctuations retrieved by our inversion only predict on average a drastic loss of energy with height, but do not directly provide information about where the energy is going to. To determine the exact fraction of the energy dissipated by radiation, thermal conduction, or changes of the ionization fraction requires a detailed study of individual events.
\subsection{Visibility and extent of magnetic canopies in temperature maps}
Up to here, we discussed the properties of chromospheric brightenings and
  the chromospheric thermal structure in the quiet Sun as a purely hydrodynamical phenomenon without considering the presence of magnetic fields. The simultaneous polarimetric observations of the 630\,nm channel of POLIS, however, also
  provide information on polarisation signals, hence photospheric magnetic
  fields. We can thus investigate the influence of the magnetic topology on the
  thermal structure in the quiet Sun as well.

The typical behaviour of the \ion{Ca}{ii} H line with its repeated change from reversal-free absorption to emission and vice versa in ``field-free'' regions, i.e., on locations that are polarisation-signal-free down to amplitudes of about 10$^{-3}$ of $I_c$, and with a persistent stationary emission on locations with photospheric polarisation signal fits to the idea of magnetic field lines as a mediator between a stationary and transient emission pattern in \ion{Ca}{ii} H. For photospheric magnetic flux concentrations, geometrical models can be constructed by setting field strength and total magnetic flux in the photosphere, and requiring (magneto-)hydrostatic pressure balance with their surroundings. This yields the classical ``expanding flux tubes'' whose diameter increases with height in the atmosphere \citep[e.g.,][SO91]{solanki+etal1991}. Now, if the magnetic fields have a profound influence on the chromospheric temperature structure, they have to show up in the temperature stratifications retrieved by the inversion, regardless of the LTE -- NLTE problem. 

Figure \ref{previous_fig} shows that this is (clearly) the case. Panel {\em a} shows the temporally averaged temperature stratifications along the slit vs.~optical depth. The also temporally averaged unsigned Stokes $V$ signal is overplotted as a {\em black line} as a reference of photospheric magnetic fields that were roughly persistent over the duration of the time series of about one hour. Already in the temporally averaged temperature, flux-tube-like features are found co-spatial to the largest values of $|V|$. To increase the visibility, and because the main effect of the magnetic fields is an increase of temperature relative to field-free surroundings, we subtracted the temporally and spatially averaged temperature stratification from the temporally averaged temperatures (panel {\em b}). In that case, temperature enhancements of +50, +100, and +150\,K over the average temperature almost exclusively occur on locations of photospheric magnetic fields. For the two strong magnetic flux concentrations at $x\sim 14.5\,$Mm and $35\,$Mm, the temperature enhancements at chromospheric layers above 500\,km given by the +50\,K contour form exactly the ``wine-glass'' shape predicted and used in many photospheric flux tube models \citep[SO91,][their Fig.~10]{pizzo+etal1993,gadun+etal2001,uitenbroek2011,khomenko+collados2012}. In both cases, the temperature enhancements yield a nearly horizontal (temperature) canopy at a height of about 620\,km that extends between 1 and 5\,Mm horizontally away from the central axis \citep[see also][]{delacruz+etal2013}. The temperature enhancements delineate vertical structures at the end of the canopies. 

The restricted horizontal extent of the temperature enhancements is better
seen in panel {\em c} that shows the temperature stratifications for one step
of the time series. Close to photospheric polarisation signals, the nearly
horizontal canopies in temperature appear again prominently, especially from
$x\sim 6$ to $16$\,Mm where three neighboring flux tubes create a continuous
temperature canopy, but in-between separated flux tubes areas of several Mm
extent with a strongly reduced temperature show up ($x\sim 0\,-7$\,Mm, $x\sim
23\,-33$\,Mm). In similar-sized areas, reversal-free \ion{Ca}{ii} H spectra
were found before \citep{rezaei+etal2008}. In total, this implies that on the
one hand magnetic canopies have a direct effect on the temperature
stratification even in a single snapshot of the solar atmosphere, and on the
other hand there are no clear indications for {\em volume-filling}
magnetic fields in the formation height of \ion{Ca}{ii} H in the temperature
map of Fig.~\ref{previous_fig}. The appearance changes a bit in the
temperature map for the large-area scan at high layers  (observation no.~1 of BE09, cf.~Fig.~\ref{corr_fig_1}).

 Figure \ref{previous_fig} clearly demonstrates that photospheric magnetic
  fields are a second acting agent for the chromospheric temperature structure
  in addition to the transient BGs. A more detailed study of the modification of the thermal structure at or close to the locations of photospheric magnetic fields is, however, beyond the scope of the present paper.
\subsection{Applicability of LTE inversion}
Because of the known problems of assuming LTE in an inversion of any chromospheric line, it might be worth to note a few examples for which purposes the LTE inversion still serves without any, or with only weak restrictions. We remark that mainly the modulus of the temperature derived from the LTE inversion has a high inaccuracy because of the missing instantaneous coupling between gas temperature and radiation field in NLTE conditions, and that the conversion from optical depth to geometrical height suffers from the lack of hydrostatical equilibrium. The latter could, however, be obtained post-facto with a presumably minor impact on the resulting synthetic spectra.

The detectability of canopies in temperature maps -- that are implicitly
assumed to correspond to the theoretically expected magnetic canopies -- does
depend only slightly on the modulus of temperature. Our results clearly show
that the signature of the canopies can be detected in the surroundings of
photospheric magnetic flux concentrations up to a few Mm
(cf.~Fig.~\ref{previous_fig}). This is  in contrast to the results of
\citet{judge+etal2010} who found a strong ``confinement'' of intensity
enhancements, and hence magnetic flux in the chromosphere.  The
  mismatch of these two results shows that a quantitative in-depth analysis of spectra, even with the limiting LTE assumptions, is crucial for tracing the topology at chromospheric layers. 

For estimating the energy of individual BGs, we used a height range, or equivalently, a wavelength region in the line wing, where NLTE effects can be assumed to play a minor role ($z<500$\,km). These results are thus rather robust and would presumably not change significantly when a different inversion method would be used. The advantage of the inversion of the \ion{Ca}{ii} H spectra now is to provide on the one hand also a quantitative guess for the corresponding properties at the atmosphere layers above, which cannot be accessed by any photospheric line, and on the other hand to provide clear information on the temporal evolution, such as the vertical propagation speed (Fig.~\ref{speed_fig}), or the exact propagation direction from combining horizontal motions in the 2D sections of the FOV (Fig.~\ref{size_fig}) with the vertical propagation speed. 

\section{Summary and conclusions \label{conclusion}}
In a LTE inversion of \ion{Ca}{ii} H spectra, we cannot retrieve on average a chromospheric temperature rise in quiet Sun (QS) regions, whereas in active regions (ARs) this is the case. The difference between QS and ARs can be directly traced back to the observed input profiles themselves, independent of the LTE assumption in the inversion. We confirm the findings of BE12 of a layer at about 150\,km height with an increased occurence of high-intensity, and hence high-temperature events near and in magnetic features. A comparison of line-core and line-wing images, and the correlation coefficient between intensities and temperature vs.~optical depth suggests that the increased skewness of the intensity histograms on magnetic locations is caused by a direct signature of magnetic fields in temperature at two different atmosphere layers, i.e., both in the chromosphere and the mid photosphere. The magnetic network is inconspicuous at the layers in-between. The appearance of the network in the mid photosphere could be caused by the ``hot wall'' effect in combination with the shift of the optical depth scale inside magnetic fields \citep{spruit1976}, whereas the appearance in the chromosphere indicates an energy deposit instead.

The energy contained in the rms fluctuations of temperature and the large difference between QS and AR profiles and temperature stratifications indicate the importance of chromospheric heating processes related to the presence of magnetic fields. The physical processes that could yield such a heating should be investigated in more detail in the future and included in the analysis of chromospheric lines. The temperature fluctuations in the QS fall short of the requirements needed to maintain a stationary chromospheric temperature rise.

The properties of individual (heating) events related to chromospheric bright grains (BGs) can be determined during all of their existence in the LTE inversion results, from the BG precursors in the photosphere to their culmination in the lower chromosphere. The mechanical energy content and the vertical energy flux in the photosphere would be sufficient to fulfill the generic chromospheric energy requirement of about 4\,kW\,m$^{-2}$, but only a fraction well below 5\,\% of the mechanical energy reaches the chromosphere, falling short by far of the generic requirement. The decrease of the internal energy density with height agrees with that expected for acoustic waves.

The topology of magnetic fields in the chromosphere is seemingly directly reflected in the temperature structure at layers above a height of 500\,km also in an LTE analysis, as indicated by the appearance of canopy structures around magnetic field concentrations both in average and in instantaneous $x-z$ temperature maps. With all of the shortcomings of the LTE approach, this quantitative analysis of large-scale data sets in a chromospheric spectral line provides a wealth of interesting features that can be used to address open questions in solar chromospheric physics of which this paper still has not covered all despite its extent.
\begin{acknowledgements}
The VTT is operated by the Kiepenheuer-Institut f\"ur Sonnenphysik (KIS) at the
Spanish Observatorio del Teide of the Instituto de Astrof\'{\i}sica de Canarias (IAC). The POLIS instrument has been a joint development of the High Altitude Observatory (Boulder, USA) and the KIS. C.B.~acknowledges partial support by the Spanish Ministry of Science and Innovation through project AYA2010--18029 and JCI-2009-04504. R.R. acknowledges financial support by the DFG grant RE 3282/1-1. We thank H.~Socas-Navarro (IAC) and T.~Carroll for helpful discussions.
\end{acknowledgements}
\bibliographystyle{aa}
\bibliography{references_luis_mod_partIII}
\Online
\begin{appendix}
\section{Effects of spectral broadening\label{appa}}
The effective spectral resolution of the POLIS spectrograph can lead to a loss in amplitude of the emission peaks by spectral smearing because of their small spectral extent. The default spectral resolution of POLIS is 220.000@400\,nm ($\equiv 1.8$\,pm), but because of the on-chip binning by a factor of two, the resolution is sampling-limited to about 3.8\,pm. Convolving an FTS \citep{kurucz+etal1984} reference spectrum with the method of \citet{allendeprieto+etal2004} and \citet{cabrera+etal2007} to match the observed average Ca profile yielded the need for a Gaussian with $\sigma \sim 1.7$\,pm. We note that the FTS spectrum contains line broadening by macroturbulent velocities because of being a spatially averaged QS spectrum.

For an estimate of the maximal spectral smearing between the observed average QS profile and the FAL-C NLTE profile -- synthesized without a macroturbulent velocity -- by instrumental effects, we first convolved the FAL-C NLTE profile as reference by a Gaussian to match the shape and width of the line blends in the Ca wing in the average observed spectrum. The {\em lower left panel} of Fig.~\ref{conv_fig} shows that this requires a value of $\sigma$ of about 3.4\,pm, which is in rough agreement with the prediction from the sampling limit. A convolution of the full spectrum with this value, as an estimate of the instrumental and velocity broadening in the atmosphere, reduces, however, the amplitude of the emission peaks of the FAL-C NLTE profile only slightly ({\em lower right panel} of Fig.~\ref{conv_fig}). To force the emission peaks in the synthetic profile to a shape roughly resembling the observed profile requires a convolution of FAL-C with a Gaussian with $\sigma = 9.4$\,pm. The line blends in the wing, however, clearly exclude such a large instrumental broadening because in the  FAL-C NLTE profile convolved with $\sigma = 9.4$\,pm ({\em purple line}) the blends are twice as broad and half as deep as in the observed spectrum ({\em lower left panel} of Fig.~\ref{conv_fig}).

\begin{figure}
\begin{minipage}{8.8cm}
\resizebox{8.8cm}{!}{\includegraphics{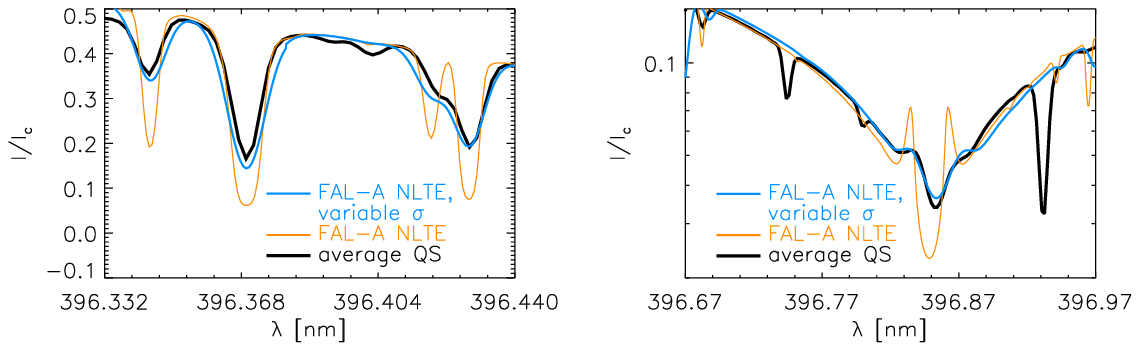}}\\
\resizebox{8.8cm}{!}{\includegraphics{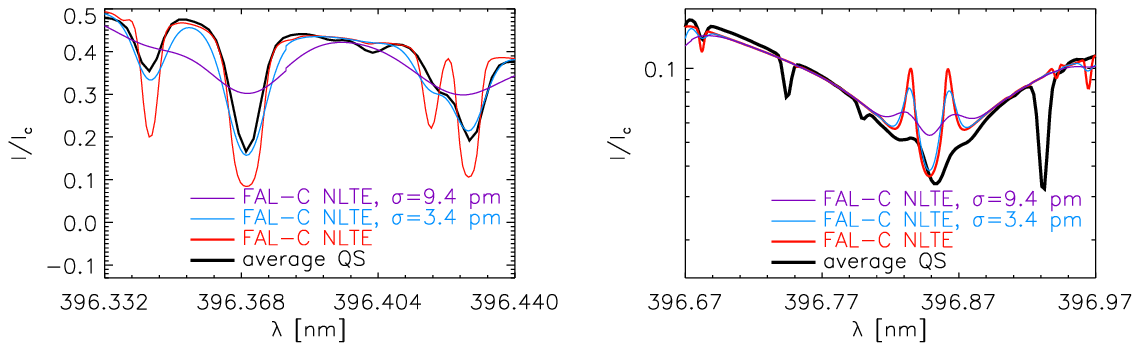}}
\end{minipage}
\caption{Convolution of the FAL NLTE profiles with different spectral PSFs. {\em Bottom}: convolution of FAL-C with Gaussians of $\sigma = 3.4$\,pm ({\em blue}) and 9.4\,pm ({\em purple}). The {\em black line} shows the observed average QS profile. {\em Left column}: line wing. {\em Right column}: line core. {\em Top}: convolution of FAL-A with a wavelength-dependent Gaussian with $\sigma$ between 3.4 and 9.4\,pm.\label{conv_fig}}
\end{figure}

We then decided to try a wavelength-dependent width of the Gaussian used as spectral point spread function (PSF). The width was set to 3.4\,pm in the line wing up to $\lambda=396.75$\,nm, gradually increasing to 9.4\,pm at $\lambda=396.85$\,nm, and reducing down to 3.4\,pm again at $\lambda=396.94$\,nm. As it seemed impossible to match the FAL-C NLTE and the observed profile even with a large broadening in the very core, we applied this variable convolution to the FAL-A NLTE profile instead. The temperature stratification of FAL-A has a weaker chromospheric temperature rise than FAL-C (Fig.~\ref{fig_temp_comp}) and the corresponding  NLTE profile differs significantly from FAL-C, especially in the residual line-core intensity (Fig.~\ref{av_profs}). The approach with the variable spectral PSF maintains the match in the line wing ({\em upper left panel} of Fig.~\ref{conv_fig}) and leads to an acceptable agreement between the emission peaks of the convolved FAL-A NLTE profile and the observed spectrum ({\em upper right panel}). 

The wavelength-dependent Gaussian spectral PSF would correspond to a microturbulent velocity $v_{mic}$ that changes from 2.6\,kms$^{-1}$ in the photosphere to 7.1\,kms$^{-1}$ in the chromosphere. The NLTE synthesis already included a height-dependent microturbulent velocity of more than 5\,kms$^{-1}$ for $\log\tau < -5.5$. Thus, the effective microturbulent velocity necessary to match the averaged observed and the FAL-A NLTE profile should be nearly twice as large as the initial value of $v_{mic}$, and has to be present already at lower layers of $\log \tau \sim -3$ to yield the correct shape of the H$_{\rm 1V}$ and H$_{\rm 1R}$ minima next to the emission peaks. We note that similar values of $v_{mic}$ would be also required to match the FTS atlas and the FAL-A NLTE profile, because the line shape in the FTS is similar to the average QS spectra shown here. If such a high value of $v_{mic}$ is reasonable for spectra at about 1$^{\prime\prime}$ spatial resolution needs a further investigation. Additionally, the increase of $v_{mic}$ at comparably low layers of $\log \tau$ also already affects the \ion{Fe}{i} line blend at 396.93\,nm strongly, broadening it significantly beyond its width in observed spectra. Doubling $v_{mic}$ therefore seems not to be a valid option to match observed and theoretical NLTE profiles, leaving a reduction of temperature as the only option \citep[cf.][]{rezaei+etal2008}.
\end{appendix}
\end{document}